\documentclass[manuscript]{acmart}

\usepackage{moresize}

\usepackage[ruled, linesnumbered, longend, lined]{algorithm2e} % For algorithms
\SetAlFnt{\fontsize{7pt}{7pt}\selectfont}
\newcommand{\algoFontSize}{tiny}

\SetCommentSty{mycommentfont}

\newcommand{\tableFontSize}{small}

\usepackage{multicol}
\usepackage{multirow}
\usepackage[inline]{enumitem}
\usepackage{bm}
\usepackage{graphicx}
\usepackage{caption}
\usepackage{array}
\usepackage{nicefrac}
\usepackage{titlecaps}
\usepackage{hyperref}

\usepackage{pifont}
\newcommand{\cmark}{\text{\ding{51}}}
\newcommand{\xmark}{\text{\ding{55}}}

\usepackage{xpatch}

\makeatletter
\newcommand\thefontsize[1]{{#1 The current font size is: \f@size pt\par}}
\xpatchcmd{\ps@firstpagestyle}{Manuscript submitted to ACM}{}{\typeout{First patch succeeded}}{\typeout{first patch failed}}
\xpatchcmd{\ps@standardpagestyle}{Manuscript submitted to ACM}{}{\typeout{Second patch succeeded}}{\typeout{Second patch failed}}    
\@ACM@manuscriptfalse% Also in titlepage
\makeatother

\renewcommand\footnotetextcopyrightpermission[1]{} % removes footnote with conference info
\setcopyright{none}
\title{An Adaptive Approach to Recoverable Mutual Exclusion}
\author{Sahil Dhoked}
\affiliation{
  \institution{The University of Texas at Dallas}
  \state{TX}
  \postcode{75080}
  \country{USA}}
\email{sahil.dhoked@utdallas.edu}
\author{Neeraj Mittal}
\affiliation{
  \institution{The University of Texas at Dallas}
  \state{TX}
  \postcode{75080}
  \country{USA}}
\email{neerajm@utdallas.edu}
\usepackage{tikz,circuitikz}
\usetikzlibrary{shapes.geometric, shapes.multipart, arrows, chains, positioning, fit, backgrounds, decorations.pathreplacing, decorations.markings, calc}

\tikzstyle{qnode} = [rectangle split, rectangle split parts=2, rectangle split horizontal, draw=black]

\newcommand{\bigO}[1]{\mathcal{O}(#1)}
\newcommand{\bigOmega}[1]{{\Omega}(#1)}
\newcommand{\bigTheta}[1]{{\Theta}(#1)}
\newcommand{\card}[1]{\bigm| #1 \bigm|\ }

\newcommand{\CAS}{\textsf{CAS}}
\newcommand{\FAS}{\textsf{FAS}}

\newcommand{\F}{F}
\newcommand{\n}{n}

\newcommand{\NCS}{\texttt{NCS}}
\newcommand{\CS}{\texttt{CS}}
\newcommand{\Recover}{\texttt{Recover}}
\newcommand{\Enter}{\texttt{Enter}}
\newcommand{\Exit}{\texttt{Exit}}
\newcommand{\segment}{segment}

\newcommand{\consequence}{consequence}

\newcommand{\responsive}{responsive}

\newcommand{\true}{\textsf{true}}
\newcommand{\false}{\textsf{false}}

\newcommand{\varstate}{state}
\newcommand{\varnode}{mine}
\newcommand{\varpred}{pred}
\newcommand{\varpath}{type}
\newcommand{\varnext}{next}
\newcommand{\varlocked}{locked}
\newcommand{\vartail}{tail}

\newcommand{\FAST}{\textsc{FAST}}
\newcommand{\SLOW}{\textsc{SLOW}}

\newcommand{\LEFT}{\textsc{Left}}
\newcommand{\RIGHT}{\textsc{Right}}

\newcommand{\theleftside}{the {\LEFT} side}
\newcommand{\therightside}{the {\RIGHT} side}

\newcommand{\PC}{PM}

\newcommand{\safe}{safe}
\newcommand{\unsafe}{unsafe}

\newcommand{\locality}{locality}

\newcommand{\nonadaptive}{non-adaptive}
\newcommand{\cadaptive}{semi-adaptive}
\newcommand{\adaptive}{adaptive}
\newcommand{\sadaptive}{super-adaptive}
\newcommand{\unbounded}{unbounded}
\newcommand{\bounded}{bounded}
\newcommand{\wbounded}{well-bounded}
\newcommand{\fadaptive}{unbounded adaptive}
\newcommand{\badaptive}{bounded-adaptive}

\newcommand{\sefficient}{\wbounded{} \sadaptive}

\newcommand{\SEfficient}{\sefficient}

\newcommand{\core}{core}
\newcommand{\splitter}{splitter}
\newcommand{\arbitrator}{arbitrator}
\newcommand{\filter}{filter}
\newcommand{\target}{target}

\newcommand{\symfilter}{\mathcal{F}} %% symbol for filter lock
\newcommand{\symcore}{\mathcal{C}} %% symbol for core lock
\newcommand{\symarbitrator}{\mathcal{A}} %% symbol for arbitrator lock

\newcommand{\weakMCS}{\textsc{WR-Lock}}

\newcommand{\base}{base}

\newcommand{\fast}{fast}
\newcommand{\slow}{slow}
\newcommand{\medium}{medium-slow}
\newcommand{\normal}{normal}

\newcommand{\mynull}{\textbf{null}}
\newcommand{\mynext}{next}
\newcommand{\mylocked}{locked}
\newcommand{\mytail}{tail}

\newcommand{\salock}{\textsc{SA-Lock}}
\newcommand{\balock}{\textsc{BA-Lock}}
\newcommand{\nalock}{\textsc{NA-Lock}}

\newcommand{\BSLock}{\salock}

\newcommand{\InNCS}{\textsc{Free}}
\newcommand{\InRecover}{\textsc{Initializing}}
\newcommand{\InEnter}{\textsc{Trying}}
\newcommand{\InCS}{\textsc{InCS}}
\newcommand{\InExit}{\textsc{Leaving}}

\newcommand{\YA}{\arbitrator} %Variable name for instance of 2-process lock
\newcommand{\sr}{\core}  %Variable name for instance of strongly recoverable lock
\newcommand{\w}{\filter}  %Variable name for instance of weakly recoverable lock
\newcommand{\x}{owner}  %Variable name for instance of selector lock
\newcommand{\s}{\mathcal{S}}  %Variable name for generic lock for adaptive algorithm
\newcommand{\fails}{k} %Variable for number of failures
\newcommand{\levels}{m} %Total number of levels of recursion
\newcommand{\threshold}{$\tau$} %Threshold for batch failures

\newcommand{\procinlock}{\ensuremath{\mathbb{P}}}
\newcommand{\failinlock}{\ensuremath{\mathbb{UF}}}
\newcommand{\Super}{\ensuremath{\Pi}}

\newcommand{\setP}[2]{\procinlock(#1, #2)}
\newcommand{\setF}[2]{\failinlock(#1, #2)}
\newcommand{\setSuper}[2]{\Super(#1, #2)}

\newcommand{\allF}{\Phi} %set of all failures

\newcommand{\sen}{sensitive}
\newcommand{\ins}{instruction}
\newcommand{\senins}{\sen{} \ins{}}

\newcommand{\switch}{switch} %Track switching of pool buffers
\newcommand{\inCounter}{in} %Number of requests generated
\newcommand{\outCounter}{out} %Number of requests satisfied
\newcommand{\pool}{pool} %Variable to store pool/buffer of nodes
\newcommand{\poolIndex}{pool\_index} %Variable to determine which pool to access
\newcommand{\nodeIndex}{node\_index} %Variable to determine which node to access from the pool

\newcommand{\JJJ}{Jayanti, Jayanti and Joshi}

\usepackage{subfiles}
\begin{document}

    \begin{abstract}
        Mutual exclusion (ME) is one of the most commonly used techniques to handle conflicts in concurrent systems. Traditionally, mutual exclusion algorithms have been designed under the assumption that a process does not fail while acquiring/releasing a lock or while executing its critical section. However, failures do occur in real life, potentially leaving the lock in an inconsistent state. This gives rise to the problem of \emph{recoverable mutual exclusion (RME)} that involves designing a mutual exclusion algorithm that can tolerate failures, while maintaining safety and liveness properties.
        
        One of the important measures of performance of any ME algorithm, including an RME algorithm, is the number of \emph{remote memory references (RMRs)} made by a process (for acquiring and releasing a lock as well as recovering the lock structure after a failure). The best known RME algorithm solves the problem for $\n$ processes in sub-logarithmic number of RMRs, given by $\bigO{\nicefrac{\log \n}{\log \log \n}}$, irrespective of the number of failures in the system. 
        
        In this work, we present a new algorithm for solving the RME problem whose RMR complexity gradually \emph{adapts} to the number of failures that have occurred in the system ``recently''. In the absence of failures, our algorithm generates only $\bigO{1}$ RMRs. Furthermore, its RMR complexity is given by $\bigO{\min\{ \sqrt{\F}, \nicefrac{\log \n}{\log \log \n} \}}$ where $\F$ is the total number of failures in the ``recent'' past.
        In addition to read and write instructions, our algorithm uses compare-and-swap (\CAS{}) and fetch-and-store (\FAS{}) hardware instructions, both of which are commonly available in most modern processors. 
        
        %Inspired by - prework
        %O(1) with swap
        
        %WHY do we need???
    \end{abstract}

\maketitle

    \section{Introduction}
    
    %% MUTUAL EXCLUSION

    One of the most commonly used techniques to handle contention in a concurrent system is to use \emph{mutual exclusion (ME)}. The mutual exclusion problem was first defined by Dijkstra more than half a century ago in~\cite{Dij:1965:CACM}. Using locks that provide mutual exclusion enables a process to execute its critical section (part of the program that involves accessing shared resources) in isolation without worrying about interference from other processes. This avoids race conditions, thereby ensuring that the system always stays in a consistent state and produces correct outcome under all scenarios.

    %% FAILURES

    Generally, algorithms for mutual exclusion are designed with the assumption that failures do not occur, especially while a process is accessing a lock or a shared resource. However, such failures can occur in the real world.
    A power outage or network failure might create an unrecoverable situation causing processes to be unable to continue. If such failures occur, traditional mutual exclusion algorithms, which are not designed to operate properly under failures, may deadlock or otherwise fail to guarantee important safety and liveness properties. 
    In many cases, such failures may have disastrous consequences.
    This gives rise to the \emph{recoverable mutual exclusion (RME) problem}. The RME problem involves designing an 
    algorithm that ensures mutual exclusion under the assumption that process failures may occur at \emph{any} point during their execution, but the system is able to recover from such failures and proceed without any adverse consequences.

    %% NVRAM
    
    Traditionally, concurrent algorithms use checkpointing and logging to tolerate failures by regularly saving relevant portion of application state to a persistent storage such as hard disk drive (HDD). Accessing a disk is orders of magnitude slower than accessing main memory. As a result, checkpointing and logging algorithms are often designed to minimize disk accesses.
    \emph{Non-volatile random-access memory (NVRAM)} memory (NVRAM) is a new class of memory technologies that combines the low latency and high bandwidth of traditional random access memory with the density, non-volatility, and economic characteristic of traditional storage media (\emph{e.g.}, HDD). Existing checkpointing and logging algorithms can be modified to use NVRAMs instead of disks to yield better performance, but, in doing so,we would not be leveraging the true power of NVRAMs \cite{NarHod:2010:ASPLOS, GolRam:2019:DC}. NVRAMs can be used to directly store implementation specific variables and, as such, have the potential for providing near-instantaneous recovery from failures. 
    
    Most of the application data can be easily recovered after failures by directly storing implementation variables on NVRAMs. However, recovery of implementation variables alone is not enough. Processor state information such as contents of program counter, CPU registers and execution stack cannot be recovered completely and need to be handled separately. Due to this reason, there is a renewed interest in developing fast and dependable algorithms for solving many important computing problems in software systems vulnerable to process failures using NVRAMs. Using innovative methods, with NVRAMs in mind, we aim to design efficient and robust fault-tolerant algorithms for solving mutual exclusion and other important concurrent problems.

    %% RELATED WORK
    
    The RME problem in the current form was formally defined a few years ago by Golab and Ramaraju in~\cite{GolRam:2016:PODC}. Several algorithms have been proposed to solve this problem~\cite{GolRam:2019:DC, GolHen:2017:PODC, JayJos:2017:DISC,JayJay+:2019:PODC}. One of the most important measures of performance of an RME algorithm is the maximum number of \emph{remote memory references (RMRs)} made by a process per critical section request in order to acquire and release the lock as well as recover the lock after a failure. Whether or not a memory reference incurs an RMR depends on the underlying memory model. The two most common memory models used to analyze the performance of an RME algorithm are \emph{cache-coherent (CC)} and \emph{distributed shared memory (DSM)} models.  
    
    RMR complexity of existing RME algorithms differs under different scenarios. For example, the RME algorithms presented in~\cite{GolRam:2019:DC} has RMR complexity of $\bigO{1}$ in the absence of failures and it grows linearly with the number of failures. As such, the RMR complexity of the algorithm may become arbitrarily large if a process fails repeatedly. On the other hand, the RME algorithm in~\cite{JayJay+:2019:PODC} has RMR complexity of $\bigO{\nicefrac{\log \n}{\log \log \n}}$, where $\n$ is the number of processes in the system, irrespective of how many failures have occurred in the system (including the case when the system has not experienced any failures).
	To our knowledge, all existing RME algorithms have worst-case RMR complexity of at least $\bigOmega{\nicefrac{\log \n}{\log \log \n}}$. 
	A more detailed description of the related work is given later in \autoref{sec:related}.

    \begin{table}
    \renewcommand{\arraystretch}{1.25}
        \caption{Comparison of known solutions to recoverable mutual exclusion problem with respect to RMR complexity under three different scenarios.}
        \begin{\tableFontSize}
            \begin{tabular}{||m{20em}|c|c|c||}
                \hline
                 \multicolumn{1}{||c|}{\multirow{2}{*}{\bf Algorithm}} & \multicolumn{3}{|c||}{\bf RMR Complexity} \\ \cline{2-4}
                 & {\bf No failures} & {\bf $\mathbf \F$ failures} & {\bf Arbitrarily large number of failures} \\
                 \hline \hline
                 
                 Golab and Ramaraju's transformation for recoverability \cite[Section~4.1]{GolRam:2019:DC} using MCS lock & $\bigO{1}$ & $\bigO{\F}$ & unbounded\\
                \hline
                
                 Golab and Ramaraju's transformation for bounding RMR complexity \cite[Section~4.2]{GolRam:2019:DC} using MCS lock & $\bigO{1}$ & $\bigO{\n}$ & $\bigO{\n}$\\
                \hline
                
                 Golab and Hendler's arbitration tree using $k$-port MCS lock$^\ast\dagger$ \cite{GolHen:2017:PODC} & $\bigO{\nicefrac{\log \n}{\log \log \n}}$ & $\bigO{\nicefrac{\log \n}{\log \log \n}}$ & $\bigO{\nicefrac{\log \n}{\log \log \n}}$\\
                \hline
                
                Jayanti and Joshi's wait-free recovery \cite{JayJos:2017:DISC} & $\bigO{\log \n}$ & $\bigO{\log \n}$ & $\bigO{\log \n}$\\
                \hline
                
                \JJJ{'s} arbitration tree using $k$-port MCS lock \cite{JayJay+:2019:PODC} & $\bigO{\nicefrac{\log n}{\log \log \n}}$ & $\bigO{\nicefrac{\log \n}{\log \log \n}}$ & 
                $\bigO{\nicefrac{\log \n}{\log \log \n}}$\\
                \hline
                
                Chan and Woelfel's infinite array based recoverable lock$\ddagger$ \cite{ChaWoe:2020:PODC} & $\bigO{1}$ & $\bigO{F}$ & unbounded\\
                \hline
                \hline
                Our algorithm [this work] & $\bigO{1}$ & $\bigO{\sqrt{\F}}$ & $\bigO{\nicefrac{\log \n}{\log \log \n}}$\\
                \hline
                \multicolumn{1}{l}{$\n$: number of processes in the system} &
                \multicolumn{3}{l}{$\ast$: It has been recently shown in \cite{JayJay+:2019:PODC} that the algorithm is prone to deadlocks} \\
                 \multicolumn{1}{l}{$\dagger$: RMR complexity measures only hold for the CC model} &
                 \multicolumn{3}{l}{$\ddagger$: RMR complexity is constant in the amortized case}
            \end{tabular}
        \end{\tableFontSize}
        \label{table:RMRComparisons}
    \end{table}

    \paragraph{Our Contributions:}
	In this work, we present an RME algorithm with the following desirable properties under both CC and DSM models. First, it has \emph{constant} RMR complexity in the absence of failures. Second, its RMR complexity grows \emph{sub-linearly} with (specifically, as \emph{square-root} of) the number of failures that have occurred in the system in the ``recent'' past. Third, it has \emph{sub-logarithmic} worst-case RMR complexity of $\bigO{\nicefrac{\log \n}{\log \log \n}}$, where $\n$ denotes the total number of processes in the system.
	We are not aware of any existing RME algorithm that satisfies all of the above three properties. 
	Table~\ref{table:RMRComparisons} compares the performance of different RME algorithms under a variety of situations.
		
	In addition to mutual exclusion and starvation freedom properties, our RME algorithm also satisfies bounded exit, bounded recovery and bounded critical section reentry properties. Roughly speaking, an RME algorithm satisfies the bounded exit property if a process is able to leave its critical section within a bounded number of its own steps unless it fails. It satisfies the bounded recovery property if a process is able to recover from a failure within a bounded number of its own steps unless it fails again. Finally, it satisfies the the critical section reentry property if, when a process $p$ fails inside its critical section, then no other process enters its critical section until $p$ has (re)entered its critical section. Finally, our RME algorithm also satisfies FCFS (first-come-first-served) property in the absence of failures. 
	
	The main idea behind our approach is to use a solution to a weaker variant of the RME problem, in which a failure may cause the mutual exclusion property to be violated temporarily albeit in a controlled manner, repeatedly as a filter to limit contention and achieve adaptability. 
	
	Our approach is general enough that it can be used to transform \emph{any} non-adaptive RME algorithm with worst-case RMR complexity of $T(\n)$ under a given memory model (CC or DSM) into an adaptive RME algorithm whose worst-case RMR complexity is still $\bigO{T(\n)}$ under the same memory model.

    \paragraph{Roadmap:}
    The rest of the text is organized as follows. We describe our system model and formally define the RME problem in \autoref{sec:model|problem}. 
    We define the weaker variant of the RME problem and its properties in \autoref{sec:weak_recoverability}.
    We present a highly efficient solution to the weaker variant of the RME problem with constant RMR complexity in \autoref{sec:weak|MCS}. 
    In \autoref{sec:framework|basic}, we present a framework to transform any given RME algorithm into a new RME algorithm that preserves the worst-case RMR complexity of the original RME algorithm but has lower RMR complexity in the absence of failures. This transformation uses a solution to the weaker variant of the RME problem as a building block. Applying this transformation recursively, we create a new transformation in \autoref{sec:framework|recursive} that preserves the worst-case RMR complexity of the original RME algorithm and whose performance degrades sub-linearly ($\sqrt{F}$) with the number of ``recent'' failures. This transformation achieves the desired RMR complexity for each of the three scenarios mentioned earlier (as shown in \autoref{table:RMRComparisons}).
    A detailed description of the related work is given in \autoref{sec:related}.
    Finally, in \autoref{sec:conclusion}, we present our conclusions and outline directions for future research.

    \section{System Model and Problem Formulation}
    \label{sec:model|problem}
    
        We follow the same model as used by Golab and Ramaraju in their work on recoverable mutual exclusion (RME) \cite{GolRam:2019:DC}. 
        
    \subsection{System model}
    
        We consider an asynchronous shared-memory system consisting of $\n$ unreliable processes labeled $p_1, p_2, \ldots, p_\n$. Shared memory is used to store variables that can be accessed by any process. Besides shared memory, each process also has its own private memory that is used to store variables that can only be accessed by that process (\emph{e.g.}, program counter, CPU registers, execution stack, \emph{etc.}). Processes can only communicate by performing read, write and read-modify-write (RMW) instructions on shared variables. Processes are not assumed to be reliable and may fail.
        
        A system execution is modeled as a sequence of process steps. In each step, some process either performs some local computation affecting only its private variables or executes one of the available instructions (read, write or RMW) on a shared variable or fails. Processes may run at arbitrary speeds and their steps may interleave arbitrarily. In any execution, between two successive steps of a process, other processes can perform an unbounded but finite number of steps.
        %% Processes are not assumed to be reliable and may fail.
    
        %%
        %% SHOULD REMOVE THE NEXT SENTENCE
        %%
        %% Using the variables in private and shared memory, processes compete with each other to acquire a lock in order to execute their respective critical sections.
        %%
        
        To avoid race conditions resulting from multiple processes trying to access the same shared resource simultaneously, processes synchronize their accesses to shared resources using a \emph{lock} that provides mutual exclusion (ME); at most one process can hold the lock at any time.
      
   \subsection{Failure model}
        We assume the \emph{crash-recover} failure model. A process may fail at any time during its execution by crashing. A crashed process recovers eventually and restarts its execution. A crashed process does not perform any steps until it has restarted. A process may fail multiple times, and multiple processes may fail concurrently.
        
        Note that, upon restarting after a failure, the state of the lock as well as the underlying application utilizing the lock needs to be restored to a proper condition. In this work, we focus only on the recovery of the internal structure of the lock. Restoring the application state to its proper condition (using logs and/or persistent memory) is assumed to be the responsibility of the programmer and is beyond the scope of this work~\cite{GolRam:2019:DC,GolHen:2017:PODC,JayJay+:2019:PODC}.

        On crashing, a process loses the contents of its private variables, including but not limited to the contents of its program counter, CPU registers and execution stack. However, the contents of the shared variables remain unaffected and are assumed to persist despite any number of failures. When a crashed process restarts, all its private variables are reset to their initial values.
        
        Processes that have crashed are difficult to distinguish from processes that are running arbitrarily slow. However, we assume that every process is live in the sense that a process that has not crashed eventually executes its next step and a process that has crashed eventually recovers. In this work, we consider a failure to be associated with a single process. If a failure causes multiple processes to crash, we treat each process crash as a separate failure. 
        
        %% In this work, we only consider benign crash failures. Processes may either crash, or run as 
        %% expected. A non-crashed process cannot perform any malicious activities.

        %Mention how the following example violates fairness
        %
        %   lock1.lock()
        %   lock2.lock()
        %   lock1.unlock()
        %   lock2.unlock()
        %

    \subsection{Process execution model}
        A process execution is modeled using two types of computations, namely \emph{non-critical section} and \emph{critical section}. A critical section refers to the part of the application program in which a process needs to access shared resources in isolation. A non-critical section refers to the remainder of the application program. 
       
        If multiple processes access and modify shared resource(s) concurrently, it may lead to race conditions which may prevent the application from working properly and may possibly have disastrous consequences. To avoid such race conditions, a lock (or a mutual exclusion algorithm) is used to enable each process to execute its critical section in isolation. At most one process can hold the lock at any time, and a process can execute its critical section only if it is holding the lock. The lock can be granted to another request only after the process (more specifically, request) holding the lock releases it after completing its critical section.
        Hereafter, we use the terms ``mutual exclusion algorithm'', ``ME algorithm'' and ``lock'' interchangeably.

    \begin{algorithm}[t]
        \begin{\algoFontSize}
            \DontPrintSemicolon
            \While {true} 
            {
                Non-Critical Section (NCS)\;
                Recover\;
                Enter\;
                Critical Section (CS)\;
                Exit\;
            }
            \caption{Process execution model}
            \label{algo_PEM}
        \end{\algoFontSize}
    \end{algorithm}

        The execution model of a process with respect to a lock is depicted in Algorithm~\ref{algo_PEM}. As shown, a process repeatedly executes the following five  \segment{s} in order: \NCS{}, \Recover{}, \Enter{}, \CS{} and \Exit{}.
        The first \segment{}, referred to as \NCS, models the steps executed by a process in which it only accesses variables outside the lock.
        The second \segment{}, referred to as \Recover, models the steps executed by a process to perform any cleanup required due to past failures and restore the internal structure of the lock to a consistent state. 
        The third \segment{}, referred to as \Enter, models the steps executed by a process to acquire the lock so that it can execute its critical section in isolation.
        The fourth \segment{}, referred to as \CS, models the steps executed by a process in the critical section where it accesses shared resources in isolation.
        Finally, the fifth \segment{}, referred to as \Exit, models the steps executed by a process to release the lock it acquired earlier in \Enter{} \segment. 
       
        We assume that, in \NCS{} \segment{}, a process does not access any part of the lock or execute any computation that could potentially cause a race condition. Moreover, in \Recover, \Enter{} and \Exit{} \segment{s}, a process accesses shared variables pertaining to the lock (and the lock only).

        A process may crash at any point during its execution, including while executing \NCS, \Recover, \Enter, \CS{} or \Exit{} \segment{}.  
        We assume that a crashed process upon restarting starts its execution from the beginning of the loop shown in Algorithm~\ref{algo_PEM}, specifically from the beginning of \NCS{} \segment{}. 
        Note that any steps executed by a process to recover the application state are not explicitly modeled here. Specifically, both \NCS{} and \CS{} \segment{s} may consist of code in the beginning to recover relevant portions of the application state. 
        
		In the rest of the text, by the phrase ``acquiring a recoverable lock,'' we mean ``executing \Recover{} and \Enter{} \segment{s} (in order) of the associated RME algorithm.'' Likewise, by the phrase ``releasing a recoverable lock,'' we mean ``executing \Exit{} \segment{} of the associated RME algorithm.''

        \begin{definition}[passage]
            A \emph{passage} of a process is defined as the sequence of steps executed by the process from when it begins executing \Recover{} \segment{} to either when it finishes executing the corresponding \Exit{} \segment{} or experiences a failure, whichever occurs first.
        \end{definition}
        
        %A process is said to be in cleanup if it is executing the \texttt{Recover} section of its first passage after it experienced a failure. 
        
        \begin{definition}[failure-free passage]
            A passage of a process is said to be \emph{failure-free} if the process has successfully executed \Recover, \Enter{} and \Exit{} \segment{s} of that passage without experiencing any failures. 
        \end{definition}
        
        \begin{definition}[super-passage]
            A \emph{super-passage} of a process is a maximal non-empty sequence of consecutive passages 
            executed by the process, where only the last passage of the process in the sequence is failure-free.
        \end{definition}
        
        A request for critical section by a process $p$ is said to be satisfied if $p$ has executed a failure-free passage for that request.

    \subsection{Problem definition}
    \label{sec:problem}
    
        A \emph{history} is a collection of steps taken by processes. 
        A process $p$ is said to be \emph{active} in a history $H$ if $H$ contains at least one step by $p$.
        We assume that every critical section is finite.
        
        \begin{definition}[fair history]
        A history $H$ is said to be \emph{fair} if 
        \begin{enumerate*}[label=(\alph*)]
            \item it is finite, or 
            \item if it is infinite and every active process in $H$ either executes infinitely many steps or stops taking steps after a failure-free passage.
        \end{enumerate*}
        \end{definition}
        
        Designing a recoverable mutual exclusion (RME) algorithm involves designing \Recover, \Enter{} and \Exit{} \segment{s} such that the following correctness properties are satisfied.  
        
        \begin{description}
        
            \item[Mutual Exclusion (ME)] For any finite history $H$, at most one process is in its \CS{} at the end of $H$. 
        
            \item[Starvation Freedom (SF)] Let $H$ be an infinite fair history in which every process crashes only a finite number of times in each super passage. Then, if a process $p$ leaves the \NCS{} \segment{} in some step of $H$, then $p$ eventually enters its \CS{} \segment{}.
            
            \item[Bounded Critical Section Reentry (BCSR)] For any history $H$, if a process $p$ crashes inside its \CS{} \segment{}, then, until $p$ has reentered its \CS{} \segment{} at least once, any subsequent execution of \Enter{} \segment{} by $p$ either completes within a bounded number of $p$'s own steps or ends with $p$ crashing.
        \end{description}
        
        Note that mutual exclusion is a safety property, and starvation freedom is a liveness property. The bounded critical section reentry is a safety as well as a liveness property. If a process fails inside its \CS{}, then a shared object or resource (\emph{e.g.}, a shared data structure) may be left in an inconsistent state. The bounded critical section reentry property allows such a process to ``fix'' the shared resource before any other process can enter its \CS{} (\emph{e.g.}, \cite{GolRam:2019:DC,GolHen:2017:PODC,JayJay+:2019:PODC}). This property assumes that the \CS{} is idempotent; i.e, the \CS{} is designed so that, in a super passage, multiple executions of the \CS{} is equivalent to one execution of the \CS{}.

        Our correctness properties are the same as those used in \cite{GolRam:2019:DC,GolHen:2017:PODC,JayJay+:2019:PODC}. We have stated them here for the sake of completeness.
        In addition to the correctness properties, it is also desirable for an RME algorithm to satisfy the following additional properties.
        
        \begin{description}
        
            \item[Bounded Exit (BE)] For any infinite history $H$, any execution of the \Exit{} \segment{} by any process $p$ either completes in a bounded number of $p$'s own steps or ends with $p$ crashing.
         
            \item[Bounded Recovery (BR)] For any infinite history $H$, any execution of \Recover{} \segment{} by process $p$ either completes in a bounded number of $p$'s own steps or ends with $p$ crashing.
        
        \end{description}
        
    \subsection{Performance measures}
    
        We measure the performance of RME algorithms in terms of the number of \emph{remote memory references (RMRs)} incurred by the algorithm during a \emph{single} passage. The definition of a remote memory reference depends on the memory model implemented by the underlying hardware architecture. In particular, we consider the two most popular shared memory models:
        
       \begin{description}
            \item[Cache Coherent (CC)]
                The CC model assumes a centralized main memory. Each process has access to the central shared memory in addition to its local cache memory. The shared variables, when needed, are cached in the local memory. These variables may be invalidated if updated by another process. Reading from an invalidated variable causes a cache miss and requires the variable value to be fetched from the main memory. Similarly, write on shared variables is performed on the main memory. Under this model, a remote memory reference occurs each time there is a fetch operation from or a write operation to the main memory.
            \item[Distributed Shared Memory (DSM)]
                The DSM model has no centralized memory. Shared variables reside on individual process nodes. These variables may be accessed by processes either via the interconnect or a local memory read, depending on where the variable resides. Under this model, a remote memory reference occurs when a process needs to perform \textit{any} operation on a variable that does not reside in its own node's memory.
        \end{description}
        
        In the rest of the text, if not explicitly specified,  
        the RMR complexity measure of an algorithm applies to \emph{both CC and DSM models}.
        
        We analyze the RMR complexity of an RME algorithm under three scenarios:
        \begin{enumerate*}[label=(\alph*)]
        \item in the absence of failures (failure free RMR complexity),
        \item in the presence of $\F$ failures (limited failures RMR complexity), and
        \item in the presence of an unbounded number of failures (arbitrary failures RMR complexity).
        \end{enumerate*}
        We identify the following desirable performance measures applicable to an RME algorithm:
        
        \begin{description}
        
            \item[\PC~1. (Constantness)] Failure free RMR complexity of the algorithm is $\bigO{1}$.
            
            \item[\PC~2. (Adaptiveness)] Limited failures RMR complexity of the algorithm is 
            \begin{enumerate}[label=(\alph*)]
                \item $\bigO{g(\F)}$, where $g(x)$ is a monotonically non-decreasing function of $x$.
                \item  $o(\F)$.
            \end{enumerate}
        \end{description}
        The RMR-complexity of an RME algorithm should be bounded if the number of failures is arbitrarily large. We define the following performance measures of boundedness for this reason.
        \begin{description}
            \item[\PC~3. (Boundedness)] Arbitrarily large number of failures RMR complexity of the algorithm is 
            \begin{enumerate}[label=(\alph*)]
                \item $\bigO{h(\n)}$, where $h(x)$ is a monotonically non-decreasing function of $x$.
                \item  $o(\log \n)$.
            \end{enumerate}
        \end{description}

    \begin{table}
    \renewcommand{\arraystretch}{1.25}
        \caption{Comparison of known solutions to recoverable mutual exclusion problem with respect to the four performance measures.}
        \begin{\tableFontSize}
            \begin{tabular}{||m{20em}|c|c|c|c|c|m{12em}||}
                \hline
                 \multicolumn{1}{||c|}{\multirow{2}{*}{Algorithm}} & \multicolumn{5}{|c|}{Performance Measure} &
                 \multicolumn{1}{|c||}{\multirow{2}{*}{Classification}}
                 \\ \cline{2-6}
                 & \PC~1 & \PC~2(a) & \PC~2(b) & \PC~3(a) & \PC~3(b) & \\
                 \hline \hline
                 
                 Golab and Ramaraju's transformation for recoverability \cite[Section~4.1]{GolRam:2016:PODC} using MCS lock & \cmark & \cmark & \xmark & \xmark & \xmark &\unbounded{} \adaptive{} \\
                \hline
                
                 Golab and Ramaraju's transformation for bounding RMR complexity \cite[Section~4.2]{GolRam:2016:PODC} using MCS lock & \cmark &  \xmark  &  \xmark  &  \cmark & \xmark &\bounded{} \cadaptive{} \\
                \hline
                
                 Golab and Hendler's arbitration tree using $k$-port MCS lock$^\ast$ \cite{GolHen:2017:PODC} &   \xmark  &  \xmark  & \xmark &   \cmark  & \cmark &\wbounded{} \nonadaptive{} \\
                \hline
                
                Jayanti and Joshi's wait-free recovery \cite{JayJos:2017:DISC} &   \xmark &  \xmark  &  \xmark  &  \cmark & \xmark &\bounded{} \nonadaptive{} \\
                \hline
                
                Jayanti and Joshi's arbitration tree using $k$-port MCS lock \cite{JayJay+:2019:PODC} &  \xmark  &  \xmark  & \xmark &  \cmark & \cmark &\wbounded{} \nonadaptive{} \\
                \hline
                \hline
                Our algorithm [this work] & \cmark &  \cmark &  \cmark & \cmark & \cmark &
                \wbounded{} \sadaptive{} \\
                \hline
                \multicolumn{4}{l}{$\ast$: it has been recently shown in \cite{JayJay+:2019:PODC} that the algorithm is prone to deadlocks}
            \end{tabular}
        \end{\tableFontSize}
        \label{table:PerformanceCharacteristics}
    \end{table}

        Note that \PC~2(a) implies \PC~1, \PC~2(b) implies \PC~2(a) and \PC~3(b) implies \PC~3(a).
        A comparison of the known RME algorithms with respect to the above performance measures \PC~1 to \PC~3  
        is shown in \autoref{table:PerformanceCharacteristics}. 
        Based on the subset of performance measures an RME algorithm satisfies, given a memory model (CC or DSM), we classify algorithms based on
        
        \begin{enumerate}
            \item Based on adaptiveness
                \begin{itemize}
                
                \item \emph{\nonadaptive} if its failure free RMR complexity is $\bigTheta{\text{arbitrary failures RMR complexity}}$.
                
                %\item \emph{\nonadaptive} if it does not satisfy \PC~1.
                
                \item \emph{\cadaptive} if it satisfies \PC~1, but not \PC~2(a).
                
                \item \emph{\adaptive} if it satisfies \PC~2(a) (hence also \PC~1).
                
                \item \emph{\sadaptive} if it satisfies \PC~2(b) (hence also \PC~2(a) and \PC~1).
                
                \end{itemize}
                
            \item Based on boundedness 
                \begin{itemize}
                
                \item \emph{\unbounded} if it does not satisfy \PC~3(a).
                               
                \item \emph{\bounded} if it satisfies \PC~3(a).
                
                \item \emph{\wbounded} if it satisfies \PC~3(b).
                
                \end{itemize}
        
        \end{enumerate}
        
        %In general, unless we use the term ``unbounded'' to describe an RME algorithm, the algorithm is assumed to have a bounded worst-case RMR complexity.
       
        As shown in \autoref{table:PerformanceCharacteristics}, all existing RME algorithm are either \nonadaptive, \cadaptive{} or \fadaptive. To our knowledge, there is no \badaptive, let alone \wbounded{} \sadaptive{} RME algorithm currently for either memory model. Note that our taxonomy may not be able to classify all possible RME algorithms (or recoverable algorithms in general), but it is sufficient for classifying and comparing existing RME algorithms. Additionally, our taxonomy 
        assumes that there is no algorithm that solves the RME problem with $\bigO{1}$ RMR complexity using only existing hardware instructions.
  
 \subsection{Synchronization primitives}
       
        We assume that, in addition to read and write instructions, the system also supports \emph{fetch-and-store (\FAS)} and \emph{compare-and-swap (\CAS{})} read-modify-write (RMW) instructions. 

        A fetch-and-store instruction takes two arguments: $address$ and $new$; it replaces the contents of a memory location ($address$) with a given value ($new$) and returns the old contents of that location.

        A compare-and-swap instruction takes three arguments: $address$, $old$ and $new$; it compares the contents of a memory location ($address$) to a given value ($old$) and, only if they are the same, modifies the contents of that location to a given new value ($new$). It returns \true{} if the contents of the location were modified and \false{} otherwise.

        Both instructions are commonly available in many modern processors such as Intel~64~\cite{Intel64Manual} and AMD64~\cite{AMD64Manual}.

 \section{Weak Recoverability}
        \label{sec:weak_recoverability}
        To design a \wbounded{} \sadaptive{} RME algorithm, we use a solution to the \emph{weaker} variant of the RME problem as a \emph{building block} in which a failure may cause the ME property to be violated albeit only temporarily and in a controlled manner. 
        We refer to this variant as the \emph{weakly recoverable mutual exclusion problem}.
        
        %%
 	    %%
 	    %% Should the properties be defined in terms of a history
 	    %%
 	    %%    
           
        To formally define how long a violation of the ME property may last, we define the notion of \emph{\consequence{} interval} of a failure.
        
        \begin{definition}[\consequence{} interval]
            The \emph{\consequence{} interval} of a failure $f$ 
            %with respect to a lock
            in a history $H$ is defined as the interval in time that starts from the onset of the failure and extends to the point in time when either all requests that were generated
            %in the lock
            before this failure occurred in $H$ have been satisfied or the last step in $H$ is performed, whichever happens earlier.
        \end{definition}
       
       Intuitively, we use the notion of \consequence{} interval of a failure to capture the maximum duration for which the impact of the failure may be felt in the system.
       
       \begin{definition}[weakly recoverable mutual exclusion]
            An algorithm is a \emph{weakly recoverable mutual exclusion algorithm} if it always satisfies the starvation freedom property and, for any finite history $H$, if two or more processes are in their critical sections simultaneously at some point in $H$, then that point overlaps with the \consequence{} interval of some failure.
            \label{def:wRME}
        \end{definition}
       
       Roughly speaking, a weakly RME algorithm satisfies the ME property as long as no failure has occurred in the ``recent'' past.
       Hereafter, to avoid confusion, we sometimes refer to the traditional recoverable mutual exclusion problem (respectively, algorithm) as defined in \autoref{sec:problem} as \emph{strongly recoverable mutual exclusion} problem (respectively, algorithm).

       The bounded exit, bounded recovery and bounded critical section reentry properties defined earlier in \autoref{sec:problem} are applicable to weakly RME problem as well.
              
       We demonstrate that it is possible to design an optimal weakly recoverable mutual exclusion algorithm using existing hardware instructions whose worst-case RMR complexity is only $\bigO{1}$ under both CC and DSM models. In contrast, the best known strongly RME algorithm has worst-case RMR complexity of $\bigOmega{\nicefrac{\log \n}{\log \log \n}}$ under both CC and DSM models. We exploit this \emph{gap} to design an RME algorithm that is adaptive and bounded. 
       To prove that our algorithm is \sefficient, we utilize some additional properties of our weakly RME algorithm.
       
       Note that, Not all failures may cause the ME property to be violated when using a weakly RME algorithm. To that end, we define the notion of \senins{} of an algorithm. 
       
       \begin{definition}[\senins{}]
           An \ins{} $\sigma$ of a weakly RME algorithm is said to be \emph{\sen} if there exists any finite history $H$ that satisfies the following conditions:
           \begin{enumerate*}[label=(\alph*)]
               \item it contains exactly one failure in which a process crashes immediately after executing said \ins{} $\sigma$ and 
               \item it does not satisfy the ME property
           \end{enumerate*};
           it is said to be \emph{non-\sen{}} otherwise.
       \end{definition}

       \begin{definition}[\unsafe{} failure]
            A failure is said to be \emph{\unsafe} with respect to a weakly RME algorithm if it involves a process crashing while (immediately before or after) performing a \senins{} with respect to the algorithm; it is said to be \emph{\safe} otherwise.
       \end{definition}
       
       Note that, by definition, every \ins{} of a strongly RME algorithm is a non-\senins{}. As a result, every failure is \safe{} with respect to a strongly RME algorithm. 
       
       The next notion limits the ``degree'' of violation (of the ME property) by a weakly RME algorithm if and when it occurs.

        \begin{definition}[\responsive{} weakly recoverable mutual exclusion]
            We say that a weakly recoverable mutual exclusion algorithm is \emph{\responsive{}} if, for all $k \geq 1$, it satisfies the following property: for any finite history $H$, if at least $k+1$ processes are in their critical sections simultaneously at some point in $H$, then that point overlaps with the \consequence{} intervals of at least $\bigOmega{k}$ (\unsafe) failures.
            \label{def:wRME|responsive}
        \end{definition}
       
        \subsection{Composite recoverable locks}
        
        The properties defined above are with respect to a \emph{single} weakly recoverable lock. In order to construct a \wbounded{} \sadaptive{} (strongly) recoverable lock with desired performance characteristics, we use multiple weakly recoverable locks. 
        We call a lock as \emph{composite} if it is employs one or more (weakly or strongly recoverable) locks. Composite locks might have several possible structures. For instance, the \Enter{} \segment{} of one lock could be contained in the \Enter{} or \CS{} \segment{} of another lock or the \CS{} \segment{} of one lock may be contained in the \NCS{} \segment{} of another lock. An example of a composite lock is a lock based on the tournament algorithm \cite{GolRam:2019:DC}.
        
        Note that, when we have multiple locks, the notions defined in the previous (sub)section, namely \consequence{} interval, \senins{} and \unsafe{} failure, become \emph{relative} to the specific lock.
        For example, a failure will have a different \consequence{} interval with respect to each lock. An \ins{} may be \sen{} with respect to one lock but non-\sen{} with respect to another.
        Thus, in a composite lock,  a failure may be \unsafe{} with respect to one or more weakly recoverable locks.

        \begin{definition}[\locality{} property]
        A composite (weakly or strongly) recoverable lock  is said to satisfy the \emph{\locality{} property} if, for any \ins{} $\sigma$, $\sigma$ is \sen{} with respect to \emph{at most one} of its component weakly recoverable locks.
        \end{definition}
        
        A composite lock whose component locks are all strongly recoverable trivially satisfies satisfies the \locality{} property.

%\subfile{Sections/BuildingBlocks}

%\subfile{Sections/GenericAlgo}

    \section{An Optimal Weakly Recoverable Lock}
    \label{sec:weak|MCS}
    
        In this section, we present a weakly recoverable lock whose RMR complexity is $\bigO{1}$ per passage for all three failure scenarios under both CC and DSM models. 
        Our lock is based on the well-known MCS queue-based (non-recoverable) lock~\cite{MelSco:1991:TrCS}. 
        The original lock did not satisfy the bounded exit property. Dvir and Taubenfeld proposed an extension to the original algorithm in~\cite{DviTau:2018:DISC} to make the \Exit{} \segment{} wait-free. 
        We extend the augmented MCS lock, which satisfies bounded-exit property, to make it weakly recoverable.
        
        \subsection{Original MCS queue based lock}
        
            Any request in the MCS mutual exclusion algorithm is represented using a node. The algorithm maintains a first-come-first-served (FCFS) queue of outstanding requests using a linked-list of their associated nodes. A node contains two fields:
            \begin{enumerate*}[label=(\alph*)]
               \item $\mynext$, which is a reference to its successor node in the queue (if any), and
               \item $\mylocked$, which is a boolean variable used by a process to spin while waiting for its turn to enter its critical section. 
            \end{enumerate*}
            The queue itself is represented using a shared variable $\mytail$ that contains reference to the last node in the queue if non-empty and \mynull{} otherwise.

        	To acquire the lock, a process first initializes its queue node by setting its $\mynext$ and $\mylocked$ fields to \mynull{} and \true{}, respectively. It then appends the node to the queue by performing an \FAS{} instruction on $\mytail$ using the reference to its own node as an argument (to the instruction). Note that the instruction returns the contents of $\mytail$ just before it is modified. 
        	If the return value is \mynull{}, then it indicates that the lock is free and the process has successfully acquired the lock. 
        	If not, then it indicates that the lock is not free and the return value is the reference to the predecessor of the process' own node in the queue. In that case, it notifies the owner of the predecessor node of its presence. To that end, it stores the reference to its own node in the $\mynext$ field of the predecessor node, thereby creating a forward link between the two nodes. It then starts spinning on the $\mylocked$ field of its own node waiting for it to be reset to \false{} by the owner of the predecessor node as part of 
        	releasing the lock.
            
            To release the lock, a process first tries to reset the $\mytail$ variable to \mynull{} (if $\mytail$ still contains the reference to this process' node) using a \CAS{} instruction. If the instruction returns true, then it implies that the queue does not contain any more outstanding requests and the lock is now free. On the other hand, if the instruction returns false, then it implies that the queue contains at least one outstanding request and its own node is guaranteed to have a successor. It then waits until the $\mynext$ field of its own node contains a valid reference (a non-null value) indicating that a link has been created between its own node and its successor. Finally, it follows this link and resets the $\mylocked$ field in its successor node to \false{}.
        
        \subsection{Adding bounded exit property}
        
            The original algorithm as described above does not satisfy the bounded-exit property since a process leaving its critical section may have to wait until a link between its own node and its successor has been created. 
            
            To achieve the bounded-exit property, the original algorithm is augmented with a mechanism that allows a leaving process to notify the process next in line acquire the lock, in case the link from the former's node to the latter's node has not been created yet, that the lock is now free. To that end, a process on leaving its critical section attempts to store a special value (\emph{e.g.}, reference to its own node) in the $\mynext$ field of its own node using a \CAS{} instruction. Likewise, a link is also created using a \CAS{} instruction instead of a simple write instruction as in the original algorithm.  Both \CAS{} instructions are designed to succeed only if the $\mynext$ field contains \mynull{} value, thereby ensuring that the $\mynext$ field can only be modified once.  
            
            Thus, if the \CAS{} instruction performed  by a process leaving its critical section returns false, then that process can conclude that the forward link has already been created and it then follows this link and resets the $\mylocked$ field of its successor node. On the other hand, if the \CAS{} instruction performed by a process trying to create the link returns false, then that process can infer that the lock is free and that it now holds the lock. 
            
            With this modification, unlike in the original algorithm,  after releasing the lock, 
            a process cannot always reuse its own node for the next request.  
            
        \subsection{Adding weak recoverability}
            
     \begin{algorithm}[t]
        \SetKw{LNot}{not}
     
        \begin{\algoFontSize}
            \begin{multicols}{2}
            
            \SetKw{Shared}{shared variables}
            \SetKw{Local}{local variables}
            \SetKw{Struct}{struct}
            \SetKw{Integer}{int}
            \SetKw{Boolean}{bool}
            \SetKw{Array}{array}
            \SetKw{Await}{await}
            
            \Struct QNode 
            \{ \\
            \Indp
            \tcc{location used for spinning while waiting to enter \CS{}} 
            $\varlocked$: boolean variable\;
            \tcc{reference to the successor node}
            $\varnext$: reference to QNode\;
            \Indm
            \}\;
            
            \BlankLine
            
            \Shared \\
            \Indp
            \tcc{reference to the last node in the queue} 
            $\vartail$: reference to QNode\;
            \tcc{state of the process with respect to the lock; in the DSM model, the $i$-{th} entry is local to process $p_i$}
            $\varstate$: \Array $[1{\dots}\n]$ of integer variables\;
            \tcc{reference to my own node; in the DSM model, the $i$-{th} entry is local to process $p_i$} 
            $\varnode[1{\dots}\n]$: \Array $[1{\dots}\n]$ of references to QNode\;
            \tcc{reference to the predecessor node; in the DSM model, the $i$-{th} entry is local to process $p_i$} 
            $\varpred[1{\dots}\n]$: \Array $[1{\dots}\n]$ of references to QNode\;
           
            \Indm
            
            \BlankLine
            
            \SetKw{Initialization}{initialization}
            \Initialization \\
            \Indp
            $\vartail$ $\leftarrow$ \mynull\tcp*[r]{queue is initially empty}
            \ForEach{$j \in \{ 1, 2, \dots, \n\}$}
            {
               $\varstate[j]$ $\leftarrow$ \InNCS\tcp*[r]{process is in \NCS{}}
            }

            \Indm
            
            \BlankLine
            
            \SetKwProg{fMCSRecover}{Function}{}{end}
            \fMCSRecover{Recover(~)}
            {
             
               \uIf{($\varstate[i]$ = \InEnter)}
               {
                  \label{line:recover:enter|if}
                  \If{($\varpred[i]$ = $\varnode[i]$)}
                  {
                     \tcc{may have failed earlier while performing \FAS{} instruction; abort the attempt}
                     \tcc{once \FAS{} step has been performed without any interruption, the two references are guaranteed to be different}
                     Exit(~)\;
                  }
                  \label{line:recover:enter|endif}
               } \ElseIf{($\varstate[i]$ = \InExit)}
               {
                  \label{line:recover:exit|if}
                  Exit(~)\tcp*[r]{finish executing \Exit{} \segment}
                 
               }
                \label{line:recover:exit|endif}
                
               \If(\tcp*[f]{initialize lock}){($\varstate[i]$ = \InNCS)}
               {
                  \label{line:recover:ncs|if}
                  $\varnode[i]$ $\leftarrow$ \mynull\tcp*[r]{reset reference to own node}
                  \label{line:recover:minenull}
                  %% $\varpred[i]$ $\leftarrow$ \mynull\;
                  $\varstate[i]$ $\leftarrow$ \InRecover\tcp*[r]{advance the state}
                 
               }
               \label{line:recover:ncs|endif}
            }
            
            \SetKwProg{fMCSEnter}{Function}{}{end}
            \fMCSEnter{Enter(~)}
            {
               \If{($\varstate[i]$ = \InRecover)}
               {
                  \label{line:enter:recover|if}
                  \If{($\varnode[i]$ = \mynull)}
                  {
                     $\varnode[i]$ $\leftarrow$ create a new node\;
                     \label{line:MCS:enter:newnode}
                  }
                  \tcc{initialize fields of my own node}
                  $\varnode[i].\varnext$ $\leftarrow$ \mynull\; %% \tcp*[r]{initialize $\mynext$ field}
                  \label{line:MCS:enter:setnext}
                  $\varnode[i].\varlocked$ $\leftarrow$ \true\; %% \tcp*[r]{initialize $\mylocked$ field}
                  \label{line:MCS:enter:setlocked}
                  \tcc{the next initilization step helps to determine if \FAS{} has been performed}
                  $\varpred[i]$ $\leftarrow$ $\varnode[i]$\;
                  \label{line:MCS:enter:setpred}
                  $\varstate[i]$ $\leftarrow$ \InEnter\tcp*[r]{advance the state}  
                  
               }
               \label{line:enter:recover|endif}

               \If{($\varstate[i]$ = \InEnter)}
               {
                  \label{line:enter:enter|if}
                  \If{($\varpred[i]$ = $\varnode[i]$)}
                  {
                     \tcc{append my own node to the queue}
                     QNode $temp$ $\leftarrow$ \FAS($\vartail$, $\varnode[i]$)\;
                     \label{line:enter:FAS}
                     \tcc{persist the result of \FAS{}}
                     $\varpred[i]$ $\leftarrow$ temp\;
                  }
                  
                  \If{($\varpred[i]$ $\neq$ \mynull)}
                  {
                     \label{line:enter:pred|if}
                     \tcc{have a predecessor; create the link}
                     \CAS($\varpred[i].\varnext$, \mynull, $\varnode[i]$)\;
                     \label{line:enter:CAS}
                     \If{($\varpred[i].\varnext$ = $\varnode[i]$)}
                     {
                        \tcc{wait for the predecessor to complete}
                     	\Await \LNot{}($\varnode[i].\varlocked$)\tcp*[r]{spin}
                     	\label{line:enter:await}
                     }
                     
                  }
                  \label{line:enter:pred|endif}
                  $\varstate[i]$ $\leftarrow$ \InCS\tcp*[r]{advance the state}
                  
                  \label{line:enter:enter|endif}
               }
            }

            \BlankLine
            
            \SetKwProg{fMCSExit}{Function}{}{end}
            \fMCSExit{Exit(~)}
            {
               
               \label{line:exit:first}
               $\varstate[i]$ $\leftarrow$ \InExit\tcp*[r]{advance the state}
               \tcc{remove my node from the queue if it has no successor}              
               \CAS($\vartail$, $\varnode[i]$, \mynull)\;
               \label{line:exit:CAS|tail}
               %% \If{\LNot{}(\CAS($\vartail$, $\varnode[i]$, \mynull))}
               %% {
                  \tcc{may have a successor; signal it to enter \CS{}}
                  \CAS($\varnode[i].\varnext$, \mynull, $\varnode[i]$)\;   
                  \label{line:exit:CAS|next}
                  
                  \If{($\varnode[i].\varnext$ $\neq$ $\varnode[i]$)}
                  {
                  	 \tcc{link already created; tell the successor to stop spinning}
                     $\varnode[i].\varnext.\varlocked$ $\leftarrow$ \false\;
                  }
                  
               %% }
               \label{line:exit:successor|stop}    
               $\varstate[i]$ $\leftarrow$ \InNCS\tcp*[r]{advance the state}   
               \label{line:exit:last}        
             
           }
            
            \end{multicols}
            \caption{Pseudocode of weakly recoverable MCS lock with wait-free exit for process $p_i$.}
            \label{algo:weak|MCS}
        \end{\algoFontSize}
     \end{algorithm}

            A pseudocode of the weakly recoverable lock is given in \autoref{algo:weak|MCS}. Our pseudocode uses the following shared variables. 
            The first variable, $\vartail$, contains the address of the last node in the queue if the queue is non-empty and \mynull{} otherwise. The next three variables, $\varstate$, $\varnode$ and $\varpred$, are arrays with one entry for each process.   
            The $i$-{th} entry of $\varstate$, denoted by $\varstate[i]$, 
            contains process $p_i$'s current state with respect to the lock (explained later). 
            The $i$-{th} entry of $\varnode$, denoted by $\varnode[i]$, contains the address of the queue node associated with process $p_i$'s most recent request.
            The $i$-{th} entry of $\varpred$, denoted by $\varpred[i]$, contains the address of the predecessor node, if any, of process $p_i$ after its node has been appended to the queue.
     
            The state of a process with respect to a lock has five possible values, namely \InNCS{}, \InRecover{}, \InEnter{}, \InCS{} and \InExit{}. 
            At the beginning, the state of a process, say $p_i$, is set to \InNCS{}.
            It is changed to \InRecover{} after $p_i$ has reset $\varnode[i]$ to \mynull{} (\autoref{line:recover:minenull}).
            It is changed to \InEnter{} after 
            \begin{enumerate*}
                \item $p_i$ has initialized $\varnode[i]$ with the address of a new node (\autoref{line:MCS:enter:newnode}),
                \item initialized the two fields of $\varnode[i]$ (lines~\ref{line:MCS:enter:setnext}~and~\ref{line:MCS:enter:setlocked}) and finally 
                \item initialized $\varpred[i]$ by setting it equal to $\varnode[i]$ (\autoref{line:MCS:enter:setpred}).
            \end{enumerate*}
            It is changed to \InCS{} after $p_i$ has acquired the lock. 
            It is changed to \InExit{} when $p_i$ starts executing the \Exit{} \segment{}. 
            Finally, it is changed to \InNCS{} again after $p_i$ finishes executing the \Exit{} \segment{}. 
            
            Our algorithm has only one \senins{}, namely the one involving the \FAS{} instruction (\autoref{line:enter:FAS}). Recall that a process uses this \ins{} to append its own node to the queue and also obtain the address of its predecessor node. 
            %% 
            %% ADDED
            If a failure occurs immediately after executing this \ins{}, then a situation may occur where the process was able to append its node to the queue, but was unable to store the address of its predecessor node to shared memory. This is because the step actually consists of two distinct steps --- performing the \FAS{} \ins{} on shared memory location \vartail{}, and storing the result of the \FAS{} \ins{} to another shared memory location, \varpred{}[i] (for persistence). 
            If a process fails immediately after executing this \ins{}, there is no easy way to recover this address (of the predecessor) based on the current knowledge of the failed process.
            The queue continues to grow beyond this node, but it would be disconnected from the previous part of the queue, thereby creating one more sub-queue. For an example, please refer to \autoref{fig:subqueue}.
            
            If a process detects that it may have failed while executing the (\FAS) \ins{}, it ``relinquishes'' its current node, informs its successor (if any) that the lock is now ``free'' using the wait-free signalling mechanism described earlier and retries acquiring the lock using a new node. This potentially creates multiple queues (or sub-queues) which may allow multiple processes to execute their critical sections concurrently, thereby violating the ME property.
            All other \ins{s} of our algorithm are non-\sen{}. We achieve that by using the following ideas.
            
                \begin{footnotesize}
        \begin{figure}
            \centering
            \begin{tikzpicture}[draw, minimum width=1cm, minimum height=0.5cm, start chain=A going right, start chain=B going right, start chain=C]
    
                    \node (P1) [qnode, on chain=A] {$p_1$};
                    \node (P2) [qnode, on chain=A] {$p_2$};
                    \node (P3) [qnode, on chain=A] {$p_3$};
                    \node (gndA) [ground, right of=P3] {};
                    \node (P4) [qnode, below of=P1, on chain=B] {$p_4$};
                    \node (P5) [qnode, on chain=B] {$p_5$};
                    \node (P6) [qnode, on chain=B] {$p_6$};
                    \node (gndB) [ground, right of=P6] {}; 
                    \node (P7) [qnode, below of=P4, on chain=C] {$p_7$};
                    \node (P8) [qnode, on chain=C] {$p_8$};
                     \node (gndC) [ground, right of=P8] {};

                    \draw[->] (P1) -- (P2);
                    \draw[dashed, ->] (P2) -- (P3);
                    %\draw[->] (P3) -- (P4);
                    \draw[->] (P4) -- (P5);
                    \draw[->] (P5) -- (P6);
                    %\draw[->] (P6) -- (P7);
                    \draw[->] (P7) -- (P8);
                    
                    \draw[-] (P3) -- (gndA);
                    \draw[-] (P6) -- (gndB);
                    \draw[-] (P8) -- (gndC);
                    
                    \node (T) [below of=P8] {$\mytail$};
                    \draw[->] (T) -- (P8);
                    
            \end{tikzpicture}
            \captionof{figure}{Processes $p_1$ \dots $p_8$ successfully append their nodes to the tail of the queue using an \FAS{} instruction. Processes $p_4$ and $p_7$ failed to capture the result value of the \FAS{} and are unable to set the next field of the nodes of $p_3$ and $p_6$. Process $p_3$ has captured the address of the node of $p_2$ and is about to set the corresponding next field on the node of $p_2$. Effectively, three sub-queues are created due to failures of $p_4$ and $p_7$.} 
            \label{fig:subqueue}
        \end{figure}
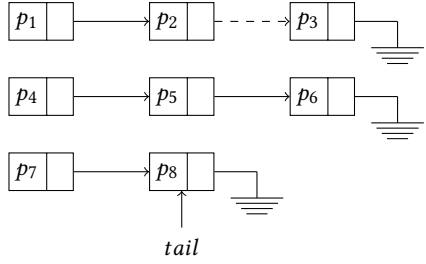
    \end{footnotesize}
            
            First, a process does not use the outcome of the \CAS{} instruction used to modify the $\mynext$ field of a node (\autoref{line:enter:CAS} and \autoref{line:exit:CAS|next}). 
            After performing the \CAS{} instruction on the $\mynext$ field, it reads the contents of the field again and determines its next step based on what it read. Note that, once initialized, the $\mynext$ field can only be modified once. This makes the two steps involving the \CAS{} instruction on the $\mynext$ field 
            as \emph{idempotent}; the effect of performing the \CAS{} instruction multiple times if interrupted due to failures is the same as performing it once. 
            
            Second, portions of \Recover{} and \Enter{} \segment{s} are enclosed in if-blocks to be executed conditionally.
            Intuitively, the guard of an if-block represents the pre-condition that needs to hold before its body can be executed. The outermost if-blocks use guards based on the current state of the process, which is advanced only at the end of the block. The inner if-blocks use guards based on other variables. Except for the if-block containing the \FAS{} instruction (which constitutes a \senins{}), all other if-blocks are idempotent and can be executed repeatedly if interrupted due to failures without any adverse impact starting from the evaluation of the guard
            (lines~\ref{line:recover:enter|if}-\ref{line:recover:enter|endif},
            lines~\ref{line:recover:exit|if}-\ref{line:recover:exit|endif},
            lines~\ref{line:recover:ncs|if}-\ref{line:recover:ncs|endif},
            lines~\ref{line:enter:recover|if}-\ref{line:enter:recover|endif} and
            lines~\ref{line:enter:pred|if}-\ref{line:enter:pred|endif}).
            Note that if the guard of an if-block does not hold, its body is not executed.
           
            Third, similar to the case of the $\mynext$ field, a process does not use the outcome of the \CAS{} instruction used to modify the $\vartail$ pointer of the queue in the \Exit{} \segment{} (\autoref{line:exit:CAS|tail}). After performing the \CAS{} instruction on the $\vartail$ pointer, irrespective of the outcome of the instruction, it blindly executes the remainder of the steps pertaining to signalling the successor node (lines~\ref{line:exit:CAS|next}-\ref{line:exit:successor|stop}). If the node has no successor, then the steps are redundant, but have no adverse impact even if the node has already been removed from the queue by an earlier \CAS{} instruction.

    \subsection{Correctness proof and complexity analysis}
          
          We refer to the algorithm described in the previous section as \weakMCS{}.
          We now prove that \weakMCS{} is a \responsive{} weakly recoverable mutual exclusion algorithm.
          
          The following proposition captures the working of the \weakMCS{} algorithm.
          
          \begin{proposition}
            \label{proposition:subqueues}
            Given a history $H$, time $t$ and $\fails \geq 0$, if at least $\fails$ processes are  in their critical sections simultaneously at time $t$, then 
            \begin{enumerate*}[label=(\alph*)]
            \item the system contains at least $\fails$ non-empty pairwise disjoint sub-queues at time $t$, and
            \item at least one node in each of the $\fails$ sub-queues is owned by a process that is in its critical section at time $t$.
            \end{enumerate*}
          \end{proposition}
          
          Note that the sub-queues may only be implicit, but can be explicitly constructed using the contents of the shared memory. We use the above proposition to argue that \weakMCS{} is \responsive. 
          
        \begin{theorem}
            \label{theorem:MCS_safety}
            Given a history $H$, time $t$ and $\fails \geq 0$, if at least $\fails + 1$ processes are  in their critical sections simultaneously at time $t$, then time $t$ overlaps with the \consequence{} interval of at least $\fails$ \unsafe{} failures.
        \end{theorem}
        \begin{proof}
           The lemma trivially holds if $\fails = 0$; therefore assume that $\fails > 0$.
           Assume that there are at least $\fails + 1$ processes  in their critical sections simultaneously at time $t$. 
           From Proposition~\ref{proposition:subqueues}, the system contains at least $\fails  + 1$ non-empty sub-queues. Only one sub-queue can have $\mytail$ pointing to its last node. Let the set of remaining sub-queues be denoted by $\mathbb{Q} = \{ Q_1, Q_2, \ldots, Q_{\fails} \}$. Note that the first node of each sub-queue is owned by a process that is in its critical section at time $t$ (From proposition~\ref{proposition:subqueues}.
           
           Consider an \emph{arbitrary} queue $Q$ from the set $\mathbb{Q}$. Let $x$ denote its last node. (Note that the last node of a sub-queue can be deduced by examining the contents of all $\varpred$ pointers in $H$.) 
           From the way the MCS algorithm works, there exists time $t' \leq t$ such that 
           $\mytail$ was pointing to $x$ at time $t'$ and some process failed while performing \FAS{} instruction on $\mytail$ at time $t'$; let that failure be denoted by $f$. 
           Let $p$ denote the process that owns a node in queue $Q$ and is in its critical section at time $t$. Clearly, $p$ generated its request before $f$ and its request is still not satisfied at time $t$. Thus, the \consequence{} interval of $f$ extends at least until time $t$.
           
           Since $Q$ was chosen arbitrarily, it follows that there exists a unique \unsafe{} failure for each of the $\fails$ sub-queues in $\mathbb{Q}$ whose \consequence{} interval extends until time $t$.
          \end{proof}
          
          \begin{theorem}
            \label{theorem:MCS_SF}
            \weakMCS{} satisfies the SF property.
          \end{theorem}
          %\begin{proof}
            %Let $H$ be an arbitrary infinite fair history. If $H$ has an infinite number of failures, then it trivially satisfies the SF property. Thus, assume that $H$ consists of a finite number of failures. Consider the execution of the system after the last failure has occurred and all processes that are crashed at that time have restarted. The system may contain multiple sub-queues at that time. Only one of these sub-queues can continue to grow, namely the one that contains the node to which $\mytail$ is pointing. All others sub-queues can only shrink.  We can view each of this sub-queue as a separate instance of the MCS lock, which is starvation free. Thus, we can conclude that every current and future request is eventually satisfied. Hence, \weakMCS{} satisfies the SF property. 
          %\end{proof}
          
          \begin{proof}
            Let $H$ be an arbitrary infinite fair history in which every process crashes only a finite number of times in each super passage. Consider an arbitrary process $p$ that has left the \NCS{} \segment{} and wants to enter its \CS{} \segment{}. Since $p$ will only fail a finite number of times in its super passage, $p$ will eventually execute the \Enter{} segment for its request. Similarly, in the \Enter{} \segment{}, process $p$ may fail in its attempt to join the queue multiple times using the \FAS{} \ins{}. However, $p$ will eventually succeed in executing the \FAS{} instruction. As a result, $p$ will join some sub-queue. 
            
            The system may contain multiple sub-queues at that time. Only one of these sub-queues can continue to grow, namely the one that contains the node to which $\mytail$ is pointing. All others sub-queues can only shrink. We can view each sub-queue as a separate instance of the MCS lock, which is starvation free. Every other process in $p$'s sub-queue will also fail only a finite number of times in their respective super passages and will eventually execute their respective \CS{} \segment{s}. Thus, $p$ will not starve and eventually enter its \CS{} \segment{}. Since $p$ was chosen arbitrarily, we can conclude that the request for every process is eventually satisfied. Hence, \weakMCS{} satisfies the SF property. 
          \end{proof}

          \begin{theorem}
          \label{theorem:MCS_BCSR}
            \weakMCS{} satisfies the BCSR property.
          \end{theorem}
          \begin{proof}
            Assume that some process $p_i$ fails while executing its \CS{} \segment, then, at the time of failure, $\varstate[i] = \InCS$. When $p_i$ restarts, after executing \NCS{} \segment{}, it executes \Recover{} \segment{} followed by \Enter{} \segment{s}. As the code inspection shows, since $state[i] = \InCS{}$, $p_i$ only evaluates a small number of if-conditions, all of which evaluate to \false, and then proceeds directly to the \CS{} \segment in a bounded number of its own steps. Hence, \weakMCS{} satisfies the BCSR property. 
          \end{proof}
          
          It follows from Theorem~\ref{theorem:MCS_safety}, Theorem~\ref{theorem:MCS_BCSR} and Theorem~\ref{theorem:MCS_SF} that
          
          \begin{theorem}
            \label{theorem:MCS_responsive}
             \weakMCS{} is a \responsive{} weakly recoverable mutual exclusion algorithm.
          \end{theorem}

          \begin{theorem}
          \weakMCS{} satisfies the BR and BE properties.
          \end{theorem}
          \begin{proof}
            As the code inspection shows, \Recover{} and \Exit{} \segment{s} do not involve any loops. Thus, a process can execute these \segment{s} within a bounded number of its own steps. Hence, \weakMCS{} satisfies the BR and BE properties.
          \end{proof}
          
          \begin{theorem}
          The RMR complexity of \Recover{}, \Enter{} and \Exit{} \segment{s} of \weakMCS{} is $\bigO{1}$ each.
          \end{theorem}
          \begin{proof}
            As the code inspection shows, \Recover{} and \Exit{} \segment{s} do not contain any loop and only contain a constant number of steps. The \Enter{} \segment{}, however has one loop at \autoref{line:enter:await} of \autoref{algo:weak|MCS}, but otherwise contain a constant number of steps. The loop involves waiting on a boolean variable until it becomes \true{} and the variable can be written to only once. In the DSM model, this variable is mapped to a location in local memory module. Hence, the RMR complexity of the \Enter{} \segment{} is also $\bigO{1}$.
          \end{proof}
   %%%%%%%%%%%%%%%%%%%%%%%%%%%%%%%%%%%%%%%%%%%%%%%%%%%     

\section{A Strongly Recoverable \SEfficient{} Lock}
\label{sec:framework}
    In this section, we describe a framework that uses other types of recoverable locks with certain properties as building blocks to construct a lock that is not only strongly recoverable but also \sefficient{} under both CC and DSM models. We describe our (\sefficient{}) lock in two steps. We first describe a basic framework to transform a \bounded{} \nonadaptive{} strongly recoverable lock to a \bounded{} \cadaptive{} strongly recoverable lock. We then extend this framework to make the lock \sadaptive{} while ensuring that it stays strongly recoverable and \bounded{}. Finally, instantiating the framework with an appropriate \nonadaptive{} lock yields the desirable \sefficient{} lock.
    
    The basic framework is based on the one used by Golab and Ramaraju in~\cite[Section~4.2]{GolRam:2019:DC} to construct a strongly recoverable lock that is \cadaptive. Specifically, in their framework, Golab and Ramaraju use two different types of strongly recoverable locks, referred to as base lock and auxiliary lock, along with two other components to build another strongly recoverable lock, referred to as target lock. The target lock constructed is \bounded{} \cadaptive{} based on the base lock that is \fadaptive{} and the auxiliary lock that is \nonadaptive. They achieve this by extending the base lock so that, upon detecting a failure, processes can abort their requests and reset the (base) lock. 
    In the presence of failures (even a single failure), the RMR complexity of the target lock is dominated by the overhead of aborting the request for the base lock and then resetting the base lock, thereby making the lock \cadaptive. In the rest of the text, we use the term ``target lock'' to refer to the (strongly recoverable) lock we want to build.
   
    \subsection{A \wbounded{} \cadaptive{} RME algorithm}
    \label{sec:framework|basic}
        
        \subsubsection{Building blocks}
        We use four different components as building blocks. 
        
        \begin{itemize}
            \item \emph{\titlecap{\filter{}} lock:} A \responsive{} weakly recoverable lock that provides mutual exclusion in the absence of failures. We use an instance of the lock proposed in \autoref{sec:weak|MCS}, which has $\bigO{1}$ RMR complexity for all three failure scenarios under both CC and DSM models.
        
            \item \emph{\titlecap{\splitter}:} Used to split processes into \emph{fast} or \emph{slow} paths. If multiple processes navigate the \splitter{} concurrently (which would happen only if an \unsafe{} failure has occurred with respect to the \filter{} lock), only one of them is allowed to take the fast path and the rest are diverted to the slow path. In other words, the \splitter{} is \emph{biased}. Intuitively, it can be viewed as a strongly recoverable \emph{try} lock. It is implemented using an atomic integer and a \CAS{} instruction, which has $\bigO{1}$ RMR complexity for all three failure scenarios under both CC and DSM models.
        
            \item \emph{\titlecap{\arbitrator{}} lock:} A \emph{dual-port} strongly recoverable lock. Each port corresponds to a side. We refer to the two sides as \emph{\LEFT} and \emph{\RIGHT}. At any time, at most one process should be allowed to attempt to acquire the lock from any side. However, \emph{any two} of the $\n$ processes can compete to acquire the lock. We use the implementation of the dual-port RME algorithm proposed by Golab and Ramaraju in~\cite[Section~3.1]{GolRam:2019:DC} (a transformation of Yang and Anderson's mutual exclusion algorithm to add recoverability), which has $\bigO{1}$ RMR complexity for all three failure scenarios under both CC and DSM models.
       
            \item \emph{\titlecap{\core{}} lock:} a (presumably \nonadaptive{}) strongly recoverable lock that assures mutual exclusion among processes taking the slow path. We may use an instance of any of the existing RME algorithms.
        \end{itemize}
        
        \subsubsection{The execution flow}
        
        In order to acquire the \target{} lock, a process proceeds as follows. It first waits to acquire the \filter{} lock. Once granted, it navigates through the \splitter{} trying to enter the fast path. If successful, it then attempts to acquire the \arbitrator{} lock from \theleftside{}. If one or more failures occur that are \unsafe{} with respect to the \filter{} lock, then multiple processes may acquire the \filter{} lock simultaneously. If this results in contention at the \splitter{}, then all but one processes are diverted to the slow path. If forced to take the slow path, the process attempts to acquire the \core{} lock. Once granted, it then waits to acquire the \arbitrator{} lock from \therightside{}. Finally, once the process has successfully acquired the \arbitrator{} lock, it is deemed to have acquired the \target{} lock as well, and is now in the \CS{} of the \target{} lock.
        
        In the absence of failures, every process takes the fast path, albeit one at a time. However, some processes do take the fast path even if their super-passage overlaps with the \consequence{} interval of an \unsafe{} failure with respect to the \filter{} lock.
        Note that at most one process can take the fast path at a time and at most one process can hold the \core{} lock at a time. Any process that takes the fast path will always attempt to acquire the \arbitrator{} lock from \theleftside{}. Any process that takes the slow path and acquires the \core{} lock will always attempt to acquire the \arbitrator{} lock from \therightside{}. Since the \core{} lock is strongly recoverable, at most one process will try to acquire the \arbitrator{} lock from each side at a time.
        
        In order to release the \target{} lock, a process simply releases its component locks in the reverse order in which it acquired them: the \arbitrator{} lock, followed by the \core{} lock (in case the process took the slow path), followed by the \splitter{} and finally the \filter{} lock.
         
        The RMR complexity of the fast path is given by the sum of the RMR complexities of the \filter{} lock, the \splitter{} and the \arbitrator{} lock. On the other hand, the RMR complexity of the slow path is given by the sum of the RMR complexities of the \filter{} lock, the \splitter{}, the \core{} lock and the \arbitrator{} lock. 
            
        For ease of exposition, we use the following terminology. Before a process is assigned a particular path, we refer to it as a \emph{\normal} process. It is classified as a \emph{\fast} process if it takes the fast path and a \emph{\slow} process otherwise. A \slow{} process becomes a \emph{\medium} process once it acquires the \core{} lock.
        
            \begin{footnotesize}
        \begin{figure}[t]
            \centering
            \begin{tikzpicture}[draw, minimum width=1cm, minimum height=1cm, scale=0.7, every node/.style={transform shape}]%,show background rectangle,inner frame sep=0mm]
                \node (W) [rectangle, draw=black, minimum height=2cm]{\titlecap{\filter} lock};
                \node (X) [diamond, draw=black, right=2cm of W]{\titlecap{\splitter}};
                \node (R) [rectangle, draw=black, minimum width=4cm, below right=3cm and 1cm of X.east]{\titlecap{\core} lock};
                \node (O) [transform shape=false,fit=(X)(R), inner sep=0mm] {};
                \node (B) [rectangle, draw=black, minimum width=3cm, minimum height=1.5cm, right=of O] {\titlecap{\arbitrator} lock};

                %Enter into W
                \draw [->] ([shift={(-3cm,0)}]W.west) -- node[anchor=south]{\Enter} (W);
                
                %Path to selector
                \draw[->] (W) -- (X);
                
                %Fast path to B
                \draw [->] (X.east) -| ([shift={(0.5cm,0)}] B.north west) node[pos=0.25,above]{fast path};
                
                %Slow path to R
                %% \draw [->] (X) -- node[anchor=west]{slow path} (X |- R.north);
                
                %Slow path to R
                \draw [->] (X.south)  |- (R.west) node[pos=0.25,right]{slow path};
                
                %% Slow path to B
                \draw [->] (R.east) -| ([shift={(0.5cm,0)}] B.south west);
                
                %Slow path within R
                %% \draw [dotted] (X |- R.north) -- (X |- R.center);
                %% \draw [dotted] (X |- R.center) -- (B |- R.center);
                %% \draw [dotted] (B |- R.center) -- (B |- R.north);
                
                %Slow path to B
                %% \draw[->] (R.north -| B) -- (B);
                
                %Boxing of all pieces
                \node [transform shape=false, draw,fit=(W)(X)(R)(B), inner sep=2mm] {};
                
                %Enter CS
                \draw [->]  (B) -- node[anchor=south]{\CS} ([shift={(3cm,0)}]B.east);
                
            \end{tikzpicture}
            \captionof{figure}{A pictorial representation of the framework.} 
            \label{fig:framework2}
        \end{figure}
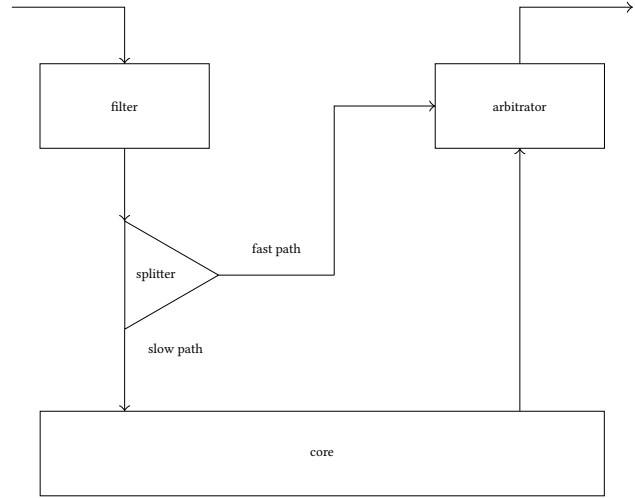
    \end{footnotesize}

     \begin{algorithm}[t]
        \begin{\algoFontSize}
            \begin{multicols}{2}
            \SetKw{Shared}{shared variables}
            \SetKw{Local}{local variables}
            \SetKw{Struct}{struct}
            \SetKw{Integer}{int}
            \SetKw{Boolean}{bool}
            \SetKw{Array}{array}
            %% \SetKw{Initialization}{initialization}
                        
            \Shared \\
            \Indp
            
            \tcc{\filter{} lock}
            $\symfilter$: $\n$-process weakly recoverabe lock\;
            \tcc{to implement \splitter{} - used to store the identifier of the process currently occupying the fast path}
            $\x$: integer variable\;
            \tcc{\core{} lock}
            $\symcore$: $\n$-process strongly recoverable lock\;
            \tcc{\arbitrator{} lock }
            $\symarbitrator$: $\n$-process dual-port strongly recoverable lock\;
            \tcc{path of the process; in the DSM model, the $i$-{th} entry is local to process $p_i$}
            $\varpath$: \Array $[1{\dots}\n]$ of boolean variables (\FAST, \SLOW)\;
            \Indm
            
			\BlankLine
			
			\SetKw{Initialization}{initialization}
			\Initialization \\
		    \Indp
			   $\x$ $\leftarrow$ 0\tcp*[r]{fast path is empty}
			
			   \ForEach{$j \in \{ 1, 2, \dots, \n\}$}
			   {
			       $\varpath[j]$ $\leftarrow$ \FAST\tcp*[r]{default path type}
			   }
		    \Indm
			
			\BlankLine
			
			\SetKw{Definitions}{definitions}
			\Definitions \\
		    \Indp
			    %if \varpath = \FAST
			   $side(\varpath) = 
                \begin{cases}
                    \LEFT{} & \text{if } \varpath = \FAST{}\\
                    \RIGHT{}, & \text{if } \varpath = \SLOW{}
                \end{cases}
               $
		    \Indm
			
			\BlankLine
			
			\SetKwProg{bfRecover}{Function}{}{end}
			\bfRecover{Recover(~)}
			{
			    
			   \tcc{In order to follow the execution model of a lock described in \autoref{sec:model|problem} 
			   (\NCS, \Recover, \Enter, \CS, \Exit{} in that order), we execute the \Recover{} \segment{}
			   of each of the recoverable locks ($\symfilter$, $\symcore$ and $\symarbitrator$) 
			   just prior to executing their respective \Enter{} \segment{s}}
			    %% \tcp{Thus, no special actions are required for recovery}
		 	}

			 \SetKwProg{bfEnter}{Function}{}{end}
			 \bfEnter{Enter(~)}
			 {
			 	$\symfilter$.Recover(~)\tcp*[r]{recover the \filter{} lock}
			    $\symfilter$.Enter(~)\tcp*[r]{acquire the \filter{} lock}

			    \If(\tcp*[f]{not yet on the slow path}){($\varpath[i]$ $\neq$ \SLOW)}
			    {
			       \CAS($\x$, 0, $i$)\tcp*[r]{attempt to take the fast path}
			       \label{line:path|fast|CAS}
			    }

			    \If(\tcp*[f]{unable to take the fast path}){($\x$ $\neq$ $i$)}
			    {
			       $\varpath[i]$ $\leftarrow$ \SLOW\tcp*[r]{committed to take the slow path}
			       \label{line:path|slow}
			       $\symcore$.Recover(~)\tcp*[r]{recover the \core{} lock}
			       $\symcore$.Enter(~)\tcp*[r]{acquire the \core{} lock}

			    }
			    
			    $\symarbitrator$.Recover($side(\varpath[i]$))\tcp*[r]{recover \arbitrator{} lock}
			    $\symarbitrator$.Enter($side(\varpath[i])$)\tcp*[r]{acquire the \arbitrator{} lock} 
			    
			 }           
			 
			 \BlankLine
			 
			 \SetKwProg{bfExit}{Function}{}{end}
			 \bfExit{Exit(~)}
			 { 
			    $\symarbitrator$.Exit($side(\varpath[i])$)\tcp*[r]{release the \arbitrator{} lock}
			    
			    %% \uIf(\tcp*[f]{took the slow path}){($\x$ $\neq$ $i$)}
			    \uIf(\tcp*[f]{took the slow path}){($\varpath[i]$ = \SLOW)}
			    {
			     	$\symcore$.Exit(~)\tcp*[r]{release the \core{} lock} 
			    	
			    } \Else(\tcp*[f]{took the fast path})
			    {
			      $\x$ $\leftarrow$ 0\tcp*[r]{the fast path is now empty}
			    }
			    $\varpath[i]$ $\leftarrow$ \FAST\tcp*[r]{reset the path type to default}
			    \label{line:path|reset}
			    $\symfilter$.Exit(~)\tcp*[r]{release the \filter{} lock}           
			 }

            \end{multicols}
            \caption{Pseudocode of the framework for designing \cadaptive{} lock for process $p_i$.}
            \label{algo:framework|semi}
        \end{\algoFontSize}
    \end{algorithm}

        A pictorial representation of the execution flow is depicted in \autoref{fig:framework2}. Note that the pictorial representation depicts the two sides of the \arbitrator{} lock as left and bottom, which actually correspond to \theleftside{} and \therightside{} of the \arbitrator{} lock respectively.
        
        The pseudocode is given in \autoref{algo:framework|semi}. The pseudocode closely follows the above description in text. A \splitter{} is implemented using an integer (shared) variable, namely $\x$. The fast path is occupied if and only if $\x$ has a non-zero value, in which case the value refers to the identifier of the process currently occupying the fast path. To take the fast path, a process attempts to store its own identifier in $\x$ using a \CAS{} instruction provided its current value is zero (\autoref{line:path|fast|CAS}). If the attempt fails, the process changes its path type to \SLOW{} (\autoref{line:path|slow}). Note that a process resets its path type from \SLOW{} to its default value of \FAST{} only after it has executed the \Exit{} \segment{} of the \core{} lock at least once without encountering any failure (\autoref{line:path|reset}). 
        
        In Golab and Ramaraju's framework, even if a process takes the fast path, it may still incur $\bigOmega{\n}$ RMR complexity in the presence of even a single failure because of the overhead of aborting requests and then resetting the base lock, which is an expensive operation. 
        In our framework, on the other hand, a process taking the fast path incurs only $\bigO{1}$ RMR complexity even with arbitrary failures because the RMR complexity of acquiring the \filter{} lock, followed by navigating the \splitter{} to take the fast path and finally acquiring the \arbitrator{} lock is only $\bigO{1}$ irrespective of the number of failures.
        %%        
        %% Also, note that a slow process is created only if an \unsafe{} failure occurs with respect to the \filter{} lock.

\subsubsection{Correctness proof and complexity analysis}
    
    We refer to the algorithm described in the previous section as \emph{\BSLock}.
    When convenient, we use $\symfilter$ and $\symcore$ to refer to the \filter{} and \core{} locks, respectively, of \BSLock.
    
    \begin{theorem}
            \BSLock{} satisfies the ME property.
            \label{theorem:algo_ME}
        \end{theorem}
        \begin{proof}
            A process enters the \CS{} \segment of  \BSLock{} after acquiring the \arbitrator{} lock from one of the sides.
            The \arbitrator{} lock satisfies the ME property as long as no more than one process attempts to acquire it from either side \LEFT{} or \RIGHT{} at any time.
            The \splitter{} ensures that, at any time, at most one process attempts to acquire the \arbitrator{} lock from \theleftside{}. 
            The \core{} lock ensures that, at any time, at most 
            one process attempts to acquire the \arbitrator{} lock from \therightside{}.
            Therefore, \BSLock{} satisfies the ME property.
        \end{proof}
        
        \begin{theorem}
             \BSLock{} satisfies the SF property.
            \label{theorem:algo_SF}
        \end{theorem}
        \begin{proof}
        We divide our proof into the following cases, based on whether failures occur or not, and if there is a failure, where does the failure occur.
        
            \begin{enumerate}[label={Case \arabic*.}, leftmargin=5\parindent]
                \item In the absence of failures:
                
                When $\fails$ processes try to acquire the lock, exactly one process acquires the \filter{} lock. This process follows the fast path owing to the \splitter{} and then acquires the ${\YA}$ lock from \theleftside{}. To release the lock, this process releases the component locks in the reverse order of acquisition. Since each of the component locks satisfy the \textit{SF} property individually, we can claim that the lock does not starve in the absence of failures.
                
                \item When failures do occur, let $p_i$ be any arbitrary failed process.
                
                \begin{enumerate}[label*=\arabic*., leftmargin=2\parindent]
                    \item If $p_i$ fails in the \Enter{} or \Exit{} section of the \filter{} lock:
                    \\Process $p_i$ will eventually restart. No process will get starved due to the SF property of the \filter{} lock (\autoref{theorem:MCS_SF}).
                    
                    \item If $p_i$ fails while navigating through the \splitter{}:
                    \\The \splitter{} does not block any process. Hence SF property can never be violated in this case.
                    
                    \item If $p_i$ fails in the \Enter{} or \Exit{} section of the $\sr$ lock:
                    \\In this case, process $p_i$ must have already acquired, but not released the \core{} lock and taken the slow path while navigating through the \splitter{}. When $p_i$ eventually restarts, it will attempt to acquire the \filter{} lock again. Since $p_i$ had initially already acquired the \filter{} lock, the BCSR property of the \filter{} lock will ensure that $p_i$ gets reentry into the critical section of the \filter{} lock. Since $p_i$ attempted to acquire the $\sr$ lock, the variable $\varpath[i]$ would have been set to $\SLOW$. Thus, process $p_i$ would retake the slow path and attempt to acquire the $\sr$ lock. Since the $\sr$ lock satisfies the SF property, and each process that fails in the $\sr$ lock, will always reacquire the $\sr$ lock, SF property will not be violated in this case.
                    
                    \item If $p_i$ fails in the \Enter{} or \Exit{} section of the $\YA$ lock from \therightside{}:
                    \\In this case, process $p_i$ must have already acquired, but not released the \core{} lock, taken the slow path while navigating through the \splitter{} and acquired but not released the $\sr$ lock. When $p_i$ eventually restarts, it will attempt to acquire the \filter{} lock again. Since $p_i$ had initially already acquired the \filter{} lock, the BCSR property of the \filter{} lock will ensure that $p_i$ gets reentry into the critical section of the \filter{} lock. Since $p_i$ attempted to acquire the $\YA$ lock from \therightside{}, the variable $\varpath[i]$ would have been set to $\SLOW$. Thus, process $p_i$ would retake the slow path and attempt to acquire the $\sr$ lock. Due to the BCSR property of the $\sr$ lock, $p_i$ will be able to successfully acquire the $\sr$ lock, and will attempt to reacquire the $\YA$ lock from \therightside{}. Since the $\YA$ lock satisfies the SF property, and each process that fails in the $\YA$ lock will always reacquire the $\YA$ lock from the same side, SF property will not be violated in this case.
                    
                    \item If $p_i$ fails in the \Enter{} or \Exit{} section of the $\YA$ lock from \theleftside{}:
                    \\In this case, process $p_i$ must have already acquired, but not released the \core{} lock and taken the fast path while navigating through the \splitter{}. At this point, the value of variable $\x$ will be set to $i$. When $p_i$ eventually restarts, it will attempt to acquire the \filter{} lock again. Since $p_i$ had initially already acquired the \filter{} lock, the BCSR property of the \filter{} lock will ensure that $p_i$ gets reentry into the critical section of the \filter{} lock. The $\x$ variable ensures $p_i$ will retake the fast path. Process $p_i$ will then continue to reacquire the $\YA$ lock from \therightside{}. Since the $\YA$ lock satisfies the SF property, and each process that fails in the $\YA$ lock will always reacquire the $\YA$ lock from the same side, SF property will not be violated in this case.
                \end{enumerate}
            \end{enumerate}
            
            Thus we have proved that SF property is not violated in any case. Hence, \BSLock{} satisfies the SF property.
        \end{proof}
        
        \begin{theorem}
            \BSLock{} satisfies the BCSR property.
            \label{theorem:algo_BCSR}
        \end{theorem}
        \begin{proof}
            If some process $p_i$ is in the \CS{} \segment{} of \BSLock{}, then it currently holds 
            the \filter{} lock and it either
            \begin{enumerate*}[label={(\alph*)}] 
                \item acquired the \arbitrator{} lock from \theleftside{} by taking the fast path or
                \item acquired the \core{} lock first and then acquired the \arbitrator{} lock from \therightside{} by taking the slow path. 
            \end{enumerate*}
            
            If $p_i$ fails in the \CS{} \segment{} of \BSLock, it determines the path it took
            by checking the $\varpath[i]$ variable and then retraces the same steps it had executed earlier. Since the \filter{} lock, the \core{} lock as well as the \arbitrator{} lock satisfy the BCSR property, $p_i$ is guaranteed to be able to acquire the requisite locks and  reenter the \CS{} \segment{} of \BSLock{} within a bounded number of its own steps. Hence, \BSLock{} satisfies the BCSR property.
        \end{proof}

        \begin{theorem}
           \BSLock{} satisfies the BE and BR properties.
           %BE is dependent on BE of core?
            \label{theorem:algo_BE_BR}
        \end{theorem}
        \begin{proof}
            The \Recover{} \segment{} of \BSLock{} is empty and hence it trivially satisfies the BR property.
            
            As part of the \Exit{} \segment{} of \BSLock{}, a
            process executes the \Exit{} \segment{} of
            the \arbitrator{} lock, optionally followed by the \Exit{} \segment{} of the \core{} lock, followed by the \Exit{} \segment{} of the \filter{} lock.
            Since each of three locks individually satisfy the BE property, it follows that \BSLock{} also satisfies the BE property.
        \end{proof}
        
        It follows from theorems~\ref{theorem:algo_ME},~\ref{theorem:algo_SF},~\ref{theorem:algo_BCSR} and~\ref{theorem:algo_BE_BR} that
        
        \begin{theorem}
            \label{theorem:framework|RME}
            \BSLock{} is a strongly recoverable lock.
        \end{theorem}

        \begin{theorem}[\BSLock{} is \bounded{} \cadaptive{}]
            The RMR complexity of \BSLock{} is $\bigO{1}$ in the absence of failures and $\bigO{T(\n)}$ with arbitrary failures, where $T(\n)$ denotes the worst-case RMR complexity of the \core{} lock for $\n$ processes.
        \end{theorem}
        \begin{proof}
            In the absence of failures, \emph{only one} process can successfully acquire the \filter{} lock (Definition~\ref{def:wRME}). This process navigates the \splitter{} in $\bigO(1)$ steps, 
            takes the fast path and acquires the \arbitrator{} lock from \theleftside{} (skipping the \core{} lock along the way). The RMR complexity of the \arbitrator{} lock is $\bigO{1}$. Thus, in this case, the RMR complexity of the \target{} lock is given by $\bigO{1}$.
            
            In the presence of failures, all $\n$ processes may be able to successfully acquire the \filter{} lock and proceed to the \splitter{}. Only one of these processes is allowed to take the fast path, which then
            attempts to acquire the \arbitrator{} lock from \theleftside{}. The remaining $(\n-1)$ processes are diverted to the slow path and have to acquire the \core{} lock and then acquire the \arbitrator{} lock from \therightside{}. Thus, in this case, the RMR complexity of the \target{} lock is given by $\bigO{T(\n)}$. 
        \end{proof}

        \begin{theorem}[\BSLock{} is \wbounded{} \cadaptive{}]
            Assume that we use \JJJ{'s} RME algorithm \cite{JayJay+:2019:PODC} to implement the \core{} lock. 
            Then, the RMR complexity of \BSLock{} is $\bigO{1}$ in the absence of failures and $\bigO{\nicefrac{\log n}{\log\log \n}}$ with arbitrary failures.
        \end{theorem}

        \begin{comment}
        
        Note that at any given time, each process that attempts to acquire the \core{} lock in a particular super-passage has attempted to acquire the \filter{} lock in the same super-passage (with respect to the \target{} lock). In fact, the processes that attempt to acquire the \core{} lock at any time is a proper subset of the number of processes that acquire the \filter{} lock. Moreover, the number of processes that attempt to acquire the \core{} lock corresponds to the number of failures in the \filter{} lock since the \filter{} lock is \responsive{} (\autoref{theorem:MCS_responsive}). To capture this notion, we define the following terms($\procinlock, \failinlock, \Super$ and formally present this relation in \autoref{theorem:failure_count|base}.
        
        \end{comment}

        In the rest of this subsection, we prove an important lemma that is crucial to establish that the lock described in the next subsection is \sadaptive.
           
        Intuitively, the set of processes that attempt to acquire the \core{} lock is strictly smaller than the set of processes that attempt to acquire the \filter{} lock. Further, the size of the former set depends on the number of \unsafe{} failures that have occurred with respect to the \filter{} lock.
        To capture this formally, we first define some notations.
        Given a lock $\ell$ and time $t$, let $\setP{\ell}{t}$ denote the set of processes that have begun executing the \Enter{} \segment{} of the lock $\ell$ before or at time $t$, but have not begun executing the corresponding \Exit{} \segment. 
        Also, let $\setF{\ell}{t}$ denote the set of all failures that are \unsafe{} with respect to the lock $\ell$ and whose \consequence{} interval extents at least until time $t$. 
        Further, if a process $p$ has a pending request with respect to the target lock at time $t$, then we use $\setSuper{p}{t}$ to denote the super-passage of $p$ with respect to the target lock at time $t$.
     
        \begin{lemma}
        \label{lemma:failure_count|base}
            Consider a time $t_{\symcore}$ such that $\card{\setP{\symcore, t_{\symcore}}} > 0$. Then there exists time $t_{\symfilter}$ with $t_{\symfilter} \leq t_{\symcore}$ such that the following properties hold.
            
            \begin{enumerate}[label=(\alph*)]
                \item $\forall p \in \setP{\symcore}{t_{\symcore}}$, $\setSuper{p}{t_{\symfilter}} = \setSuper{p}{t_{\symcore}}$,
                
                \item $\setP{\symcore}{t_{\symcore}} \subsetneq \setP{\symfilter}{t_{\symfilter}}$, and
                
                \item $\card{\setF{\symfilter}{t_{\symfilter}}} \geq \card{\setP{\symcore}{t_{\symcore}}}$.
            \end{enumerate}
        \end{lemma}
        \begin{proof}
            None of the processes in the set $\setP{\symcore}{t_{\symcore}}$ was able to take the fast path while navigating the \splitter{}. Let $q$ be the \emph{last} process in $\setP{\symcore}{t_{\symcore}}$ to read  the contents of the variable $\x$ and $t$ denote the time when it performed the read step.  Clearly, $t \leq t_{\symcore}$. Furthermore, let $r$ denote the process whose identifier was stored in $\x$ when $q$ read its contents. 
            We set $t_{\symfilter}$ to $t$. We now prove each property one-by-one.
        
            \begin{enumerate}[label=(\roman*)]
            
                \item Consider an arbitrary process $s \in \setP{\symcore, t_{\symcore}}$. Assume, by the way of contradiction, that $\setSuper{s}{t_{\symfilter}} \neq \setSuper{s}{t_{\symcore}}$. This means that process $s$ generated a new request after time $t_{\symfilter}$. Since $s \in \setP{\symcore}{t_{\symcore}}$, process $s$ read the contents of the variable $\x$ some time after $t_{\symfilter}$ but before $t_{\symcore}$. This contradicts our choice of $t_{\symfilter}$.  In other words, $\setSuper{s}{t_{\symfilter}} = \setSuper{s}{t_{\symcore}}$. Since $s$ was chosen arbitrarily, it follows that for each $p \in \setP{\symcore}{t_{\symcore}}$, $\setSuper{r}{t_{\symfilter}} = \setSuper{p}{t_{\symcore}}$. Thus the property (a) holds.
    
                \item Due to the arrangement of the locks, each process in the set $\setP{\symcore}{t_{\symcore}}$ holds the lock $\symfilter$ at time $t_{\symfilter}$, Moreover, process $r$ also holds the lock $\symfilter$ at time $t_{\symfilter}$. In other words, $\setP{\symcore}{t_{\symcore}} \subseteq \setP{\symfilter}{t_{\symfilter}}$, $r \in \setP{\symfilter}{t_{\symfilter}}$ and $r \not\in \setP{\symcore}{t_{\symcore}}$.  Thus the property (b) holds.

                \item Let $\card{\setP{\symcore}{t_{\symcore}}} = \fails$. Thus, using property (b), we can conclude that $\card{\setP{\symfilter}{t_{\symfilter}}} \geq \fails + 1$. It follows from theorem~\ref{theorem:MCS_safety} that there exist at least $\fails$ failures that are \unsafe{} relative to the lock $\symfilter$ and whose \consequence{} interval overlaps with time $t_{\symfilter}$. We have $\card{\setF{\symfilter}{t_{\symfilter}}} \geq k = \card{\setP{\symcore}{t_{\symcore}}}$. Thus the property (c) holds.
                
            \end{enumerate}
            
            This establishes the result.
        \end{proof}

               %%%%%%%%%%%%%%%%%%%%%%%%%%%%%%%%%%%%%%%%%%

        \subsection{A \sefficient{} RME algorithm}
        \label{sec:framework|recursive}
        \subsubsection{The main idea}
        
        We use the \emph{gap} between the (known) worst-case RMR complexity of implementing a weakly recoverable lock and  
        that of implementing a strongly recoverable lock to achieve our goal.
    
        The main idea is to \emph{recursively} transform the \core{} lock using instances of our \cadaptive{} lock. We transform the \core{} lock repeatedly upto a height $\levels{}$ that is equal to the RMR complexity of the non-adaptive strongly recoverable lock under arbitrary failures. The strongly recoverable lock now becomes the base case of the recursion. For ease of exposition, we refer to the \core{} lock in the base case as the \emph{\base{} lock}.
        
        Let \nalock{} be a \bounded{} (presumably \nonadaptive{}) strongly recoverable lock.
        whose worst-case RMR complexity is $\bigO{T(\n)}$ for $\n$ processes. 
        Let \salock{} denote an instance of the \cadaptive{} lock described in \autoref{sec:framework|basic}. And, finally, let \balock{} denote the \bounded{} \sadaptive{} lock that we wish to construct (\nalock{} is the base lock and \balock{} is the target lock). 
        The idea is to create $\levels = T(\n)$ levels of \salock{} such that the \core{} lock component of the \salock{}
        at each level is built using another instance of \salock{} for up to $\levels-1$ levels and using an instance of \nalock{} at the base level (level $\levels$). Let \salock$[i]$ denote the instance of \salock{} at level $i$.  Formally,

        \begin{align*}
           \balock & ~=~ \salock[1] \\
           \salock[i].\text{\core} & ~=~ \salock[i+1] \qquad \forall i \in  \{1, 2, \dots, \levels-1\} \\
           \salock[\levels].\text{\core} & ~=~ \nalock
        \end{align*}
    
\begin{figure}[t]
        \scalebox{0.27}{
            \centering
            \begin{tikzpicture}[draw, 
            					minimum width=7.5cm, 
            					minimum height=3.75cm, 
            					every node/.style={transform shape, font=\HUGE},
            					triangle/.style = {regular polygon, regular polygon sides=3},
            					border rotated/.style = {shape border rotate=270},
            					decoration={markings, mark=at position 1 with {\arrow[scale=4,black]{>}}}]

               %% Level 1
               
               \node (W1) [rectangle, draw=black] {\filter$_1$};
               \node (B1) [rectangle, draw=black, right=10cm of W1] {\arbitrator$_1$};
               \node (X1) [triangle,  draw=black, border rotated, minimum size=1cm, below=7.5cm of W1.center, anchor=west] {\splitter$_1$};
               \node (IA1) [minimum size=0cm, below=2.5cm of X1.corner 3] {};
               \node (IB1) [minimum size=0cm, right=5cm of X1.corner 1] {};
               
               %% Box all shapes at level 1
               \node (Box1) [transform shape=false, dotted, draw,fit=(W1)(B1)(IA1.center), inner sep=0mm] {};
               
               %% Draw arrows within level 1
               \draw[postaction={decorate}] (W1) -- (X1.corner 2);
               \draw[-] (X1.corner 1) -- (IB1.center) node[pos=0.5, above, yshift=-0.75cm]{fast path};
               \draw[postaction={decorate}] (IB1.center) |- (B1.west);
               		       
               %% Level 2
               \node (W2) [rectangle, draw=black, below=1cm of IA1] {\filter$_2$};
               \node (B2) [rectangle, draw=black, right=10cm of W2] {\arbitrator$_2$};
               \node (X2) [triangle,  draw=black, border rotated, minimum size=1cm, below=7.5cm of W2.center, anchor=west] {\splitter$_2$};
               \node (IA2) [minimum size=0cm, below=2.5cm of X2.corner 3] {};
               \node (IB2) [minimum size=0cm, right=5cm of X2.corner 1] {};
              
               %% Box all shapes at level 2
               \node (Box2) [transform shape=false, dotted, draw,fit=(W2)(B2)(IA2.center), inner sep=0mm] {};
              
               %% Draw arrows within level 2
               \draw[postaction={decorate}] (W2) -- (X2.corner 2);
               \draw[-] (X2.corner 1) -- (IB2.center) node[pos=0.5, above, yshift=-0.75cm]{fast path};
               \draw[postaction={decorate}] (IB2.center) |- (B2.west);

               %% Level k
               \node (Wk) [rectangle, draw=black, below=5.0cm of IA2] {\filter$_\levels$};
               \node (Bk) [rectangle, draw=black, right=10cm of Wk] {\arbitrator$_\levels$};
               \node (Xk) [triangle,  draw=black, border rotated, minimum size=1cm, below=7.5cm of Wk.center, anchor=west] {\splitter$_\levels$};
               \node (IAk) [minimum size=0cm, below=2.5cm of Xk.corner 3] {};
               \node (IBk) [minimum size=0cm, right=5cm of Xk.corner 1] {};
                       
               %% Box all shapes at level k
               \node (Boxk)  [transform shape=false, dotted, draw, fit=(Wk)(Bk)(IAk.center), inner sep=0mm] {};
   
   			   %% Draw arrows at level k
   			  
               \draw[postaction={decorate}] (Wk) -- (Xk.corner 2);
               \draw[-] (Xk.corner 1) -- (IBk.center) node[pos=0.5, above, yshift=-0.75cm]{fast path};
               \draw[postaction={decorate}] (IBk.center) |- (Bk.west);
          
          	   %% Draw base case
		       \node (WkBk) [transform shape=false, fit=(Wk)(Bk), inner sep=0mm] {};
        	   \node (Base) [rectangle, draw=black, below=1.0cm of Boxk, minimum width=25cm] {\base{} lock};

          	   %% Create invisible nodes between level 2 and level k
      
           	   \node (EA1) [minimum size=0cm, above=3.0cm of Wk] {};
          	   \node (EA2) [minimum size=0cm, above=3.0cm of Bk] {};
          	   \node (EB1) [minimum size=0cm, below=1.0cm of EA1] {};
          	   \node (EB2) [minimum size=0cm, below=1.0cm of EA2] {};

        	   %% Draw slow paths  	   
          	   \draw[postaction={decorate}] (X1.corner 3) -- (W2) node[pos=0.25, right, xshift=-1.5cm]{slow path};
          	   \draw[-] (X2.corner 3) -- (EA1.center) node[pos=0.25, right, xshift=-1.5cm]{slow path};
          	   \draw[postaction={decorate}] (Xk.corner 3) -- ([shift={(3.75cm,0)}]Base.north west) node[pos=0.25, right, xshift=-1.5cm]{slow path};
          	   
          	   %% Draw ellipses between level 2 and level k
          	   
          	   \draw[-, dotted] (EA1.center) -- (EB1.center);
          	   \draw[-, dotted] (EA2.center) -- (EB2.center);
          	   \draw[postaction={decorate}] (EB1.center) -- (Wk);

          	   %% Draw return paths 
          	   \draw[postaction={decorate}] (B2) -- (B1);  
          	   \draw[-] (Bk) -- (EB2.center);
          	   \draw[postaction={decorate}] (EA2.center) -- (B2);  
          	   \draw[postaction={decorate}] ([shift={(-3.75cm,0)}]Base.north east) -- (Bk);   
          	   
          	   \draw[postaction={decorate}] ([shift={(-5cm,2.5cm)}] W1.north) -| (W1.north);     
          	   \draw[postaction={decorate}] (B1.north) |- ([shift={(5cm,2.5cm)}] B1.north);

          	   %% Label each level
          	   \node [rectangle, left=of Box1] {Level 1}; 
          	   \node [rectangle, left=of Box2] {Level 2};   
          	   \node [rectangle, left=of Boxk] {Level $\levels$};    
          	   \node [rectangle, left=of Base] {Base Case};      
                
            \end{tikzpicture}
            
          }  
            \captionof{figure}{A pictorial representation of the recursive framework.} 
            \label{fig:framework2|recursive}
       
        \end{figure}
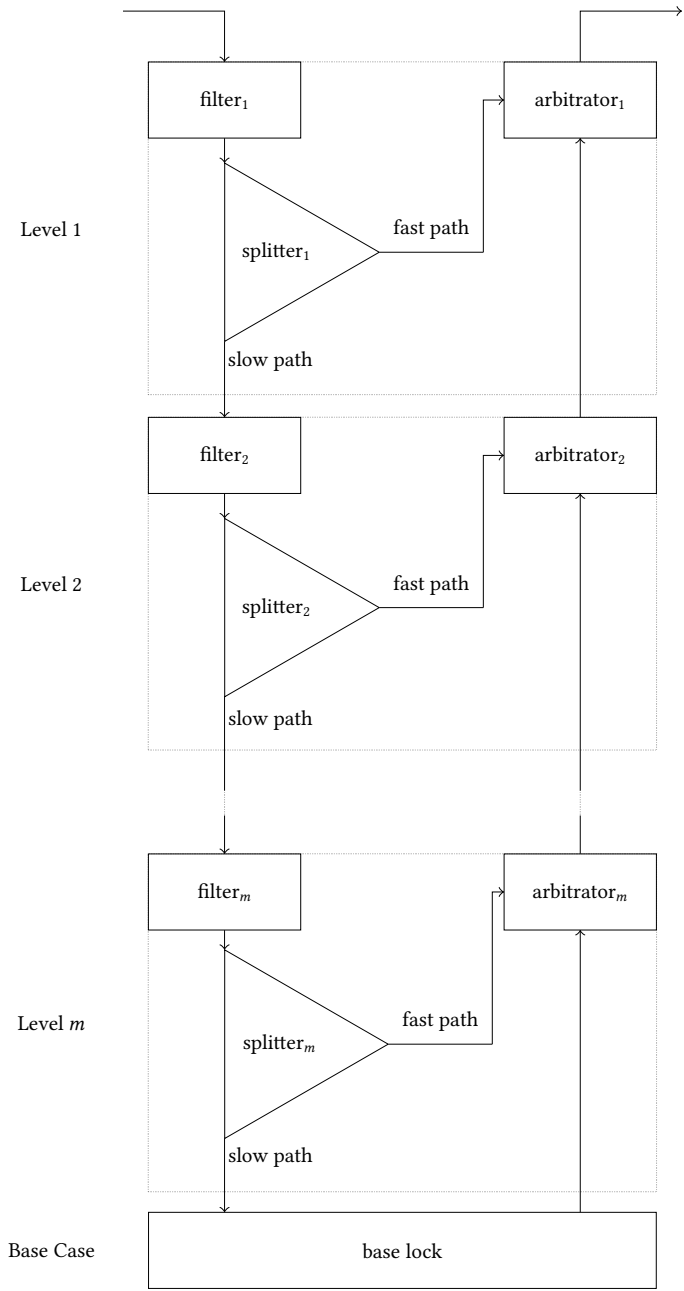

        A pictorial representation of the execution flow of the recursive framework is depicted in \autoref{fig:framework2|recursive}. 
            
        In order to acquire the \target{} lock, a process starts at the first level as a \normal{} process and waits to acquire the \filter{} lock at level 1.
        It stays on track to become a \fast{} process until an \unsafe{} failure occurs with respect to the \filter{} lock at the first level as a result of which multiple processes may be granted the (\filter{}) lock simultaneously. All of these processes then compete to enter the fast path by navigating through the \splitter. The \splitter{} allows only one process to take the fast path at a time, and the rest are diverted to take the slow path. 
        Note that a \slow{} process is created at the first level only if an \unsafe{} failure occurs with respect to the \filter{} lock at the first level. 
        All \slow{} processes at the first level then move to the second level as \normal{} processes. If no further failure occurs, then no \slow{} process is created at the second level and all processes leave this level one-by-one as \fast{} processes with respect to this level. Thus, only $\bigO{1}$ RMR complexity is \emph{added} to the passages of all the affected processes until the impact of the first failure has subsided. However, if one or more \slow{} processes are created at the second level, then we can infer that a \emph{new} \unsafe{} failure must have occurred with respect to the \filter{} lock at the second level. All these \slow{} processes at the second level then move to the third level as \normal{} processes, and so on and so forth. 
        At each level, a \slow{} process, upon either acquiring the \base{} lock or returning from the adjacent higher level (whichever case applies), becomes a \medium{} process. Irrespective of whether a process is classified as \fast{} or \medium{}, it next waits to acquire the level-specific \arbitrator{} lock. Once granted, it either returns to the adjacent lower level or, if at the initial level, is deemed to have successfully acquired the \target{} lock.
         
        Note that in our algorithm, at least $\fails$ \unsafe{} failures are required at any level to force $\fails$ processes to be ``escalated'' to the next level. Each level except for the last one would add only $\bigO{1}$ RMR complexity to the passages of these process, thus making the \target{} lock \adaptive{} under limited failures. There is no further ``escalation'' of \slow{} processes at the base level and a \bounded{} (\nonadaptive{}) strongly recoverable lock is used to manage all \slow{} processes at that point, thus bounding its RMR complexity under arbitrary failures as well. 
        
        As before, in order to release the \target{} lock, a process releases its components locks in the reverse order in which it acquired them.
         
        To prove that our \target{} lock is \sefficient{}, we utilize two properties of our framework, namely, our weakly recoverable lock is \responsive{}, and our \target{} lock, which is a composite lock, satisfies the \locality{} property.

    \subsubsection{Correctness proof and complexity analysis}

    Let $\symfilter_i$ and $\symcore_i$ denote the instances of the \filter{} and \core{} locks, respectively, at level $i$ for $i = 1, 2, \ldots, \levels$.

    \begin{theorem}
    \label{theorem:balock_safe}
        For each $i$ with $1 \leq i \leq \levels$, \salock[i] is a strongly recoverable lock.
    \end{theorem}
    \begin{proof}
        The proof is by backward induction on the level number of \salock{} starting from level $\levels$.
        \begin{itemize}[label=$\square$]
        
        \item \underline{Base case (\salock[$\levels$] is a strongly recoverable lock).} 
        Note that \salock[$\levels$] = \nalock{}. By construction, \nalock{} is a \bounded{} \nonadaptive{} strongly recoverable lock. Thus, \salock[$\levels$] is a strongly recoverable lock.
    
        \item \underline{Induction hypothesis (\salock[$i+1$] is a strongly recoverable lock for some $i$ with $1 \leq i < \levels$).} 
        
        \underline{Inductive step (\salock[$i$] is also a strongly recoverable lock).}
        Note that \salock[$i$] is an instance of our \cadaptive{} lock described in \autoref{sec:framework|basic} with 
        \salock[$i+1$] as its \core{} lock. By induction hypothesis, \salock[$i+1$] is a strongly recoverable lock. It follows from  \autoref{theorem:framework|RME} that \salock[$i$] is also a strongly recoverable lock.
        \end{itemize}
        
        \indent
        Thus, by induction, we can conclude that \salock[$i$] is a strongly recoverable lock for each $ i = 1, 2, \dots, \levels$.
    \end{proof}

    By construction, \balock{} = \salock[$1$]. Therefore,
    
    \begin{theorem}
    \balock{} is a strongly recoverable lock. 
    \end{theorem}
    
    Using induction similar to the one used in Theorem~\ref{theorem:balock_safe}, we can show that
    
    \begin{theorem}
    \balock{} satisfies the BCSR, BR and BE properties.
    \end{theorem}
    
    To analyze the RMR complexity of a passage, we first prove certain results.

    \begin{theorem}
        \balock{} satisfies the \locality{} property.
    \end{theorem}
    \begin{proof}
        \balock{} uses three types of locks, namely \filter{}, \arbitrator{} and \base{}; only \filter{} lock is weakly recoverable.  There is one instance of the \filter{} lock at each level. By construction, the \Enter{} \segment{s} of any two instances of the \filter{} lock do not overlap. The only \senins{} of the \filter{} lock is the \FAS{} instruction in its \Enter{} \segment{}. Therefore, \balock{} satisfies the \locality{} property.
    \end{proof}
    
    By the construction of our recursive framework, we have
    
    \begin{proposition}
        \label{proposition:coreisnextfilter}
        For each $i$ and time $t$ with $1 \leq i < \levels$, 
        $\setP{\symcore_{i}}{t} = \setP{\salock{[i+1]}}{t} = \setP{\symfilter_{i+1}}{t}$.
    \end{proposition}

    Note that the set of processes that attempt to acquire the \filter{} lock at any level becomes 
    \emph{progressively smaller} as the level number increases. Furthermore, the number of processes that are escalated to the next level depends on the number of \unsafe{} failures experienced by the \filter{} lock at the current level. This is captured by the next lemma.

    \begin{lemma}
    \label{lemma:level_j_failures}
    Consider a process $p$, time $t$ and level $x$, where $1 \leq x \leq \levels$,
    such that process 
    $p \in \setP{\symfilter_x}{t}$.
    Then there exist $x$ times $t_1, t_2, \ldots, t_{x-1}, t_x$ with $t_1 \leq t_2 \leq \cdots \leq t_{x-1} \leq t_x = t$ such that the following properties hold.
    For each $i$ with $1 \leq i < x$, we have
        \begin{enumerate}[label=(\alph*)]
            \item $\setSuper{p}{t_i} = \setSuper{p}{t}$, 
            \item $\setP{\symfilter_{i}}{t_{i}} \supsetneq \setP{\symfilter_{i+1}}{t_{i+1}}$, and
            \item $\card{\setF{\symfilter_{i}}{t_{i}}} \geq \card{\setP{\symfilter_{i+1}}{t_{i+1}}}$.
        \end{enumerate}
    \end{lemma}
    \begin{proof}
        The proof is by backward induction on $i$ starting from $x-1$. In order to prove our results, we use the following auxiliary properties, which are part of the induction statement. For each $i$ with $1 \leq i < x$, we have,
        \begin{enumerate}[label = (\alph*), start=4]
            \item $\card{\setP{\symfilter_i}{t_i}} > 0$, and
            \item $p \in \setP{\symfilter_i}{t_i}$.
        \end{enumerate}
    
    We are now ready to prove the result. 
    
        \begin{itemize}[label=$\square$]
        
        \item \underline{Base case (properties (a)-(e) hold for $i = x - 1$).}
        By definition, $t_x = t$. By assumption, $p \in \setP{\symfilter_x}{t_x}$. By applying  Proposition~\ref{proposition:coreisnextfilter}, we obtain that $p \in \setP{\symcore_{x-1}}{t_x}$ thereby implying that $\card{\setP{\symcore_{x-1}}{t_x}} > 0$.
        We can now apply Lemma~\ref{lemma:failure_count|base} once to infer that there exists time, say $t_{x-1}$ with $t_{x-1} < t_x$, such that the following properties hold.
        \begin{enumerate}[label=(\roman*)]
            
            \item $\setSuper{p}{t_{x-1}} = \setSuper{p}{t_x}$, which, in turn, implies that $\setSuper{p}{t_{x-1}} = \setSuper{p}{t}$ because $t_x = t$ (property (a)).
            
            \item $\setP{\symfilter_{x-1}}{t_{x-1}} \supsetneq \setP{\symcore_{x-1}}{t_x}$, which, in turn, implies that $\setP{\symfilter_{x-1}}{t_{x-1}} \supsetneq \setP{\symfilter_x}{t_x}$ because 
            $\setP{\symcore_{x-1}}{t_x} =  \setP{\symfilter_x}{t_x}$ (property (b)).
            
            \item $\card{\setF{\symfilter_{x-1}}{t_{x-1}}} \geq \card{\setP{\symcore_{x-1}}{t_{x}}}$, which, in turn, implies that $\card{\setF{\symfilter_{x-1}}{t_{x-1}}} \geq \card{\setP{\symfilter_{x}}{t_{x}}}$ (property (c)). 
            
            \item $\card{\setP{\symfilter_{x-1}}{t_{x-1}}} > 0$ because  $\setP{\symfilter_{x-1}}{t_{x-1}} \supsetneq \setP{\symfilter_x}{t_x} \supseteq \{ p \}$ (property (d)).
            
            \item $p \in \setP{\symfilter_{x-1}}{t_{x-1}}$ because $\setP{\symfilter_{x-1}}{t_{x-1}} \supsetneq \setP{\symfilter_x}{t_x} \supseteq \{ p \}$ (property (e)).

        \end{enumerate}

        \item \underline{Induction hypothesis (assume that the properties (a)-(e) hold for some $i$  with $1 < i < x$).}
        
        \underline{Inductive step (properties (a)-(e) also hold for $i-1$).} 
        Note that, by induction hypothesis, $\card{\setP{\symfilter_i}{t_i}} > 0$. Thus, We can now apply Lemma~\ref{lemma:failure_count|base} once to infer that there exists time, say $t_{i-1}$ with $t_{i-1} < t_i$, such that the following properties hold.
    
            \begin{enumerate}[label=(\roman*)]
            
            \item $\setSuper{p}{t_{i-1}} = \setSuper{p}{t_i}$, which, in turn, implies that $\setSuper{p}{t_{i-1}} = \setSuper{p}{t}$ (property (a)).
            
            \item $\setP{\symfilter_{i-1}}{t_{i-1}} \supsetneq \setP{\symcore_{i-1}}{t_i}$, which, in turn, implies that $\setP{\symfilter_{i-1}}{t_{i-1}} \supsetneq \setP{\symfilter_i}{t_i}$ because 
            $\setP{\symcore_{i-1}}{t_i} =  \setP{\symfilter_i}{t_i}$ (property (b)).
            
            \item $\card{\setF{\symfilter_{i-1}}{t_{i-1}}} \geq \card{\setP{\symcore_{i-1}}{t_{i}}}$, which, in turn, implies that $\card{\setF{\symfilter_{i-1}}{t_{i-1}}} \geq \card{\setP{\symfilter_i}{t_i}}$ (property (c)). 
            
            \item $\card{\setP{\symfilter_{i-1}}{t_{i-1}}} > 0$ because  $\setP{\symfilter_{i-1}}{t_{i-1}} \supsetneq \setP{\symfilter_i}{t_i} \supseteq \{ p \}$ (property (d)).
            
            \item $p \in \setP{\symfilter_{i-1}}{t_{i-1}}$ because $\setP{\symfilter_{i-1}}{t_{i-1}} \supsetneq \setP{\symfilter_i}{t_i} \supseteq \{ p \}$ (property (e)).

        \end{enumerate}
        
        \end{itemize}

        This establishes the lemma.
    \end{proof}
    
    The next corollary \emph{quantifies} the number of processes that must be present at \emph{each} 
    of the lower levels for some process to be escalated to a certain level.

    \begin{corollary}
        \label{corollary:count_processes_i}
         Consider a process $p$, time $t$ and level $x$, where $1 \leq x \leq \levels$,
    such that process 
    $p \in \setP{\symfilter_x}{t}$.
    Let times $t_1, t_2, \ldots, t_{x-1}, t_x$ be as given by 
    Lemma~\ref{lemma:level_j_failures}. Then, for each $i$ with $1 \leq i < x$, 
     $\card{\setP{\symfilter_i}{t_i}} \;\geq\; x-i+1$.
    \end{corollary}
    
    The next corollary \emph{quantifies} the number of \unsafe{} failures that must occur with respect to the \filter{} lock at each of the lower levels  for some process to be escalated to a certain level.

    \begin{corollary}
        \label{corollary:count_failures_i}
         Consider a process $p$, time $t$ and level $x$, where $1 \leq x \leq \levels$,
    such that process 
    $p \in \setP{\symfilter_x}{t}$.
    Let times $t_1, t_2, \ldots, t_{x-1}, t_x$ be as given by 
    Lemma~\ref{lemma:level_j_failures}. Then, for each $i$ with $1 \leq i < x$, 
    $\card{\setF{\symfilter_i}{t_i}} \;\geq\; x-i$.
    \end{corollary}

    For the rest of this section, unless otherwise stated, assume that super-passage of a process and \consequence{} interval of a failure are defined \emph{relative to the \target{} lock}.
    
     \begin{theorem}
     Suppose a process $p$ advances to level $x$ at some time $t$ during its super-passage, where $1 \leq x \leq \levels$. Then, there exist at least $\nicefrac{x (x-1)}{2}$ failures whose \consequence{} interval overlaps with the super-passage of the process $p$.
    \end{theorem}
    \begin{proof}
        Let $t_1, t_2, \dots, t_x$ be the times as given by Lemma~\ref{lemma:level_j_failures}. 
        Since \balock{} satisfies the \locality{} property, the set of failures that are \unsafe{} with respect to one instance of its \filter{} lock is \emph{disjoint} from the set of failures that are \unsafe{} with respect to another instance of its \filter{} lock. 
        Formally,
        \begin{align*}
            \forall i, j: 1 \leq i, j \leq x \text{ and } i \neq j: \setF{\symfilter_{i}}{t_i} \cap \setF{\symfilter_{j}}{t_j} = \emptyset && \text{(pairwise disjoint property)}
        \end{align*}
        
        Let $\Super$ be the super-passage of $p$ at time $t$. 
        From the property (a) of Lemma~\ref{lemma:level_j_failures}, $\Super = \setSuper{p}{t_1} = \setSuper{p}{t_2} = \ldots = \setSuper{p}{t_x}$. In other words, $p$ is executing the same super-passage during the period $[t_1,t_x]$. 
        
        Let $\allF$ denote the set of all failures whose \consequence{} interval overlaps with the super-passage $\Super$. Note that the \consequence{} interval of any failure with respect to the \target{} lock contains 
        the \consequence{} interval of that failure with respect to any instance of its \filter{} lock. This is because all pending requests for that instance of the \filter{} lock  are also pending requests for the \target{} lock. 
         Thus, $\forall i: 1 \leq i < x: \setF{\symfilter_i}{t_i} \subseteq \allF$. This in turn implies that
         \begin{align*}
         \bigcup_{i=1}^{x-1} \setF{\symfilter_i}{t_i} \; \subseteq  \; \allF && \text{(containment property)}
         \end{align*}
         
         We have
        \begin{align*}
         \card{\allF} & \geq \quad  \card{\bigcup_{i=1}^{x-1} \setF{\symfilter_i}{t_i}} && \text{(using containment property)} \\
                      & =  \quad \sum_{i=1}^{x-1}  \card{\setF{\symfilter_i}{t_i}} && \text{(using pairwise disjoint property)}\\
                      & =  \quad \sum_{i=1}^{x-1} (x-i) && \text{(using Corollary~\ref{corollary:count_failures_i})}\\
                      & =  \quad (x-1) + \cdots + 2 + 1 && \text{(expanding the sum)} \\
                      & =  \quad \frac{x(x-1)}{2} && \text{(algebra)}
        \end{align*}

        This establishes the result.
    \end{proof}
    
    \begin{theorem}[\balock{} is \bounded{} \sadaptive{}]
    If a super-passage of a process overlaps with the \consequence{} interval  of at most $\fails$ failures, then the RMR complexity of any passage in that super-passage is given by $O(\min\{\sqrt{\fails}, T(\n)\})$, where $T(\n)$ denotes the RMR complexity of the \base{} lock \nalock{} for $\n$ processes.
    \end{theorem}

    \begin{theorem}[\balock{} is \wbounded{} \sadaptive{}]
    Assume that we use an instance of \JJJ{} algorithm~\cite{JayJay+:2019:PODC} to implement 
    the \base{} lock \nalock. 
    If a super-passage of a process overlaps with the \consequence{} interval
     of at most $\fails$ failures then the RMR complexity of any passage in that super-passage is given by $O(\min\{\sqrt{\fails}, \nicefrac{\log \n}{\log \log \n}\})$. 
    \end{theorem}
   
   %%%%%%%%%%%%%%%%%%%%%%%%%%%%%%%%%%%%%%%%%%%%%%%%%

%% \subfile{Sections/AdaptiveApproach}

   \section{Related Work}
   \label{sec:related}
    
    Bohannon \emph{et al.}~\cite{BohLie+:1995:SPDP,BohLie+:1996:SPDP} were the first ones to investigate the RME problem. However, their system model is different from the one assumed in this work. Specifically, in their system model, at least one process is reliable while other processes may be unreliable. Once an unreliable process fails, it never restarts. The reliable process is responsible for continuously monitoring the health of all other processes, and, upon detecting
    that an unreliable process has failed during its passage, it performs recovery by ``fixing'' the lock. The two RME algorithms differ in the way they implement the lock; the one in~\cite{BohLie+:1995:SPDP} uses test-and-set instruction whereas the one in~\cite{BohLie+:1996:SPDP} uses MCS queue-based algorithm.
    
    Golab and Ramaraju formally defined the RME problem in~\cite{GolRam:2016:PODC}. 
    %%
    %% Their formalization, especially the system model, has served as the basis for the 
    %% subsequent work in this area, including this work. 
    %%
    We use the same system model as in their work.
    In~\cite{GolRam:2016:PODC}, Golab and Ramaraju also presented four different RME algorithms---a 2-process RME algorithm  and three $\n$-process RME algorithms. The first algorithm is based on Yang and Anderson's lock \cite{YanAnd:1995:DC}, and is used as a building block to design an $n$-process RME algorithm. Both RME algorithms use only read, write and comparison-based primitives. The worst-case RMR complexity of the 2-process algorithm is $\bigO{1}$ whereas that of the resultant $\n$-process algorithm is $\bigO{\log \n}$. Both RME algorithms have optimal RMR complexity because, as 
    shown in~\cite{AttHen+:2008:STOC, AndKim:2002:DC, YanAnd:1995:DC}, any mutual exclusion algorithm that uses only read, write and comparison-based primitives has worst-case RMR complexity of $\bigOmega{\log \n}$.  The remaining two algorithms are \fadaptive{} (with $f(x) = x$) and \cadaptive{} (with $g(x) = x$), respectively (where $f$ and $g$ are as per the definitions of adaptivity and boundedness respectively from \autoref{sec:model|problem}).
    
     Later, Golab and Hendler \cite{GolHen:2017:PODC} proposed an RME algorithm with sub-logarithmic RMR complexity of $\bigO{\nicefrac{\log \n}{\log \log \n}}$ under the CC model using MCS queue based lock~\cite{MelSco:1991:TrCS} as a building block. Note that MCS uses \FAS{} instruction, which is \emph{not} a comparison-based RMW instruction, and thus the result does not violate the previously mentioned lower bound. Their algorithm does not satisfy the bounded exit property. Moreover, it has been shown to be vulnerable to starvation~\cite{JayJay+:2019:PODC}.

    Ramaraju showed in~\cite{Ram:2015:Thesis} that it is possible to design an RME algorithm with $\bigO{1}$ RMR complexity provided the hardware provides a special RMW instruction to swap the contents of two arbitrary locations in memory atomically. Unfortunately,  at present, no hardware supports such an instruction to our knowledge.
    
    In~\cite{JayJos:2017:DISC}, Jayanti and Joshi presented an RME algorithm with $\bigO{\log{n}}$ RMR complexity.  Their algorithm satisfies bounded (wait-free) exit and FCFS (first-come-first-served) properties.
    
    In~\cite{JayJay+:2019:PODC}, \JJJ{}  proposed an RME algorithm 
    that has sub-logarithmic RMR complexity of $\bigO{\nicefrac{\log n}{\log \log n}}$. To our knowledge, this is the best known RME algorithm  as far as the worst-case RMR complexity is concerned that also satisfies bounded recovery and bounded exit properties.
    
   Using a weaker version of starvation freedom, Chan and Woelfel \cite{ChaWoe:2020:PODC} present a novel solution to the RME problem that incurs a constant number of RMRs in the amortized case, but its worst case RMR complexity may be unbounded.
   
   %However, the worst case RMR complexity of a passage in their algorithm depends on the number of failures, which may grow unboundedly. Moreover, as acknowledged by the authors, their algorithm does not satisfy the bounded (stronger) starvation freedom property.  In other words, it is possible for a (slow) process to be starved indefinitely even though every process only crashes finitely many times during its super passage. Lastly, their algorithm uses an infinite array, which makes it unsuitable even for those languages that implement their own garbage collector (\emph{e.g.}, Java). It is not clear how it can be modified to use only an array of bounded size while maintaining constant RMR amortized complexity.
    
    In~\cite{GolHen:2018:PODC}, Golab and Hendler proposed an RME algorithm under the assumption of system-wide failure (all processes fail and restart) with $\bigO{1}$ RMR complexity.

    \section{Conclusion, Discussions and Future Work}
    \label{sec:conclusion}
    
        %Impossibility of O(1) RMR algorithm
        %Node reusabililty
        %Experimental results
        
        %In our work, we have used Golab and Ramaraju's failure model~\cite{GolRam:2016:PODC} in which a failed process restarts from the beginning. An alternate failure model is to assume that a failed process restarts from the point where it crashed~\cite{JayJos:2017:DISC}. In the alternate model, instead of analyzing the RMR complexity of a passage, the RMR complexity of a super-passage is analyzed. Our RME algorithm can be easily adapted to the alternate model. If we use \JJJ{'s} RME algorithm~\cite{JayJay+:2019:PODC} to implement the \base{} lock of the  recursive framework, then the RMR complexity of a super-passage of a process  is given by $\bigO{\F_0 + \min \{ \sqrt{\F}, \nicefrac{\log \n}{\log \log \n} \}}$, where $\F_0$ denotes the number of times the process fails while executing its super-passage.
        
        %The RMR complexity of the super-passage of a process $p_i$ 
        
        \subsection{Batch failures}
            %In case of system wide failures, if we set $F = n$, it would imply that the RMR complexity of our algorithm would immediately jump to its worst case. However, that would not happen. The worst case RMR complexity result ($\bigO{\min \{ \sqrt{\F}, \nicefrac{\log \n}{\log \log \n} \}}$) is achieved when failures occur in a particular pattern, and such a pattern cannot be achieved by system wide failures. In future works, we plan to analyze the effect of system wide failures, and more generally batch failures, on our RME algorithm.

            %Introduction
            %Definitions and nomenclature
            %Theorem, proof and corollary
            %Conclusion 

            Our complexity analysis in this work considers every process failure as a unique individual failure. A system-wide failure, in which all $n$ processes fail simultaneously, is perceived as $n$ separate individual failures. So, even with a single system-wide failure, the current analysis will yield an RMR complexity of $\bigO{\nicefrac{\log n}{\log \log n}}$ per passage. However, our analysis can be extended to also incorporate batch failures. 
            
            A batch failure is a failure event where multiple processes fail simultaneously. A certain threshold \threshold{} can be established to determine if a set of failures are all individual failures or can be grouped in a single batch failure. Note that system-wide failures is a special case of batch failures.
            
            Formally, given a history $H$ and a time $t$, if the number of failures that occurred at time $t$ in $H$ is greater than the threshold \threshold{}, we call the set of failures that occurred at time $t$ as a batch failure. The consequence interval of a batch failure is the union of the consequence intervals of its component individual failures.
            
            Note that the RMR-complexity of the recursive framework depends primarily on the degree of violation of the ME property by the \filter{} lock at any level. In the worst case, a batch failure may cause every single process to ``escalate'' by one level. Intuitively, this should yield an RMR-complexity of $F_b + \sqrt{F}$, where $F_b$ is the total number of ``recent'' batch failures and $F$ is the total number of ``recent'' failures that are not part of any batch failure. We will now state this result formally% and prove it rigorously.
            
            \begin{theorem}
                Suppose a process $p$ advances to some level $x$ during its super-passage, where $1 \leq x \leq \levels$. Suppose the super-passage of $p$ overlaps with at most $u$ batch failures. Then, the super-passage of $p$ overlaps with at least $(x-u)^2$ individual failures that are not a part of any batch failure.
            \end{theorem}
            %\begin{proof}
            %    Consider some level $x' < x$.
            %\end{proof}
            
            \begin{corollary}
                Assuming that we use an instance of \JJJ{} algorithm~\cite{JayJay+:2019:PODC} to implement the \base{} lock. If a super-passage of a process overlaps with the \consequence{} interval of at most $F_b$ batch failures and $F$ individual failures that are not a part of any batch failure, then the RMR complexity of any passage in that super-passage is given by $O(\min\{F_b + \sqrt{F}, \nicefrac{\log \n}{\log \log \n}\})$.
            \end{corollary}
            
            Using this enhanced analysis, we have shown that the RMR complexity of a passage depends linearly on the number of batch failures that have occurred “recently”.
            
        \subsection{Memory Reclamation}
            A failure may prevent an MCS-queue node from being reused. Even if the owner of the node has finished executing a failure-free passage, other processes may still be accessing the node. Due to potential failures, we cannot easily determine when it would be safe to reclaim a node. For this reason, a separate memory reclamation algorithm is required to determine when it is safe to reuse a node. We use a technique similar to epoch-based memory reclamation to bound the space complexity of our RME algorithm by $\bigO{n^2 \nicefrac{\log n}{\log \log n}}$.
            %Ref to epoch based reclamation

     \begin{algorithm}
        %\begin{\algoFontSize}
            \begin{multicols}{2}
            \DontPrintSemicolon
            
            \KwData{
                \begin{itemize}[leftmargin=*]
                    \item \textbf{Shared variables:}
                    \begin{itemize}[leftmargin=*]
                        \item \pool: 3-D array $[1{\dots}\n][0,1][1{\dots}2n]$ of nodes
                        \item \inCounter: array $[1{\dots}\n]$ of integer variables
                        \item \outCounter: array $[1{\dots}\n]$ of integer variables
                        \item \switch: array $[1{\dots}\n]$ of integer variables
                        \item mode: array $[1{\dots}\n]$ of integer variables
                        \item index: array $[1{\dots}\n]$ of integer variables
                        \item snapshot: 2-D array $[1{\dots}\n][1{\dots}n]$ of integer variables
                        \item \poolIndex: array $[1{\dots}\n]$ of integer variables
                        \item confirm\_\poolIndex: array $[1{\dots}\n]$ of integer variables
                    \end{itemize}
                    \item \textbf{Local variables:}
                    \begin{itemize}[leftmargin=*]
                        \item index: integer
                        \item i: integer
                        \item \nodeIndex: integer
                    \end{itemize}
                \end{itemize}
            }
            
            \BlankLine
            
            \SetKw{Initialization}{initialization:}
            \Initialization \\
            \Indp
            \ForEach{$j \in \{ 1, 2, \dots, \n\}$}
            {
                \switch[j] $\gets$ \textsc{Completed}\;
                mode[j] $\gets$ \textsc{Scan}\;
            }
            \Indm
            
            \BlankLine
            
            \SetKwProg{newnode}{Function}{:}{end}
            \newnode{new node()}
            {
                \If{(\inCounter[i] = \outCounter[i])}
                {
                    Epoch()\;
                    \inCounter[i]++\;
                }
                \nodeIndex $\gets$ \outCounter[i] $\bmod 2n$\;
                \Return{\pool[i][\poolIndex[i]][\nodeIndex]}
            }
            
            \BlankLine
            
            \SetKwProg{retirenode}{Function}{:}{end}
            \retirenode{retire node()}
            {
                \If{(\inCounter[i] $\neq$ \outCounter[i])}{\outCounter[i]++\;}
            }
            
            \SetKwProg{Epoch}{Function}{:}{end}
            \Epoch{Epoch()}
            {
                index $\gets$ index[i]\;
                \If{(\switch[i] = \textsc{Completed})}
                {
                    \If{(mode[i] = \textsc{Scan})}
                    {
                        snapshot[i][index] $\gets$ \inCounter[index]\;
                        \uIf{(index < n)}
                        {
                            index[i]++\;
                        }
                        \uElse
                        {
                            mode[i] $\gets$ \textsc{Wait}\;
                        }
                    }
                    \If{(mode[i] = \textsc{Wait})}
                    {
                        \While{(snapshot[i][index] > \outCounter[index])}
                        {
                            \;
                        }
                        \uIf{(index > 1)}
                        {
                            index[i]-{}-\;
                        }
                        \uElse
                        {
                            \switch[i] $\gets$ \textsc{Started}\;
                        }
                    }
                }
                \If{(\switch[i] = \textsc{Started})}
                {
                    \If{(\poolIndex[i] == confirm\_\poolIndex[i])}
                    {
                        \poolIndex[i] $\gets$ 1 - \poolIndex[i]\;
                    }
                    \switch[i] $\gets$ \textsc{InProgress}\;
                }
                \If{(\switch[i] = \textsc{InProgress})}
                {
                    \If{(\poolIndex[i] $\neq$ confirm\_\poolIndex[i])}
                    {
                        confirm\_\poolIndex[i] $\gets$ \poolIndex[i]\;
                    }
                    mode[i] $\gets$ \textsc{Scan}\;
                    \switch[i] $\gets$ \textsc{Completed}\;
                }
            }
            \caption{Memory Reclamation}
            \label{algo:memrec}
            \end{multicols}
        %\end{\algoFontSize}
    \end{algorithm}

            Our memory reclamation algorithm is implemented using two methods:
            \begin{enumerate*}
                \item \textit{new node}
                \item \textit{retire node}
            \end{enumerate*}.
            This algorithm assumes that a process will only use one node at a time.
            The \textit{new node} method needs to be called whenever the \filter{} lock wants to use a new node. Similarly, the \textit{retire node} method needs to be called whenever a process is done using the node. The algorithm is designed in such a way that multiple executions of the \textit{new node} method will return the same node until the \textit{retire node} method is called. This helps us to accommodate for failure scenarios where the \textit{new node} method returns a node but the process is unable to capture the return value of the method.

            A pseudocode of the memory reclamation algorithm is presented in \autoref{algo:memrec}. Each process $p_i$ maintains two pools (active and reserve) of $2n$ nodes each (\pool$[i][0,1][1,\dots,2n]$).
            The main idea of the algorithm is to wait for \textit{old} requests of other processes to be satisfied before assigning new nodes. This helps a process $p_i$ to guarantee that no other process has any reference to its \textit{old} nodes.
            The algorithm maintains two counters:
            \begin{enumerate*}
                \item \inCounter[i]
                \item \outCounter[i]
            \end{enumerate*}, for every process $p_i$. The counter \inCounter[i] counts the number of (logical) nodes allocated to process $p_i$ and the counter \outCounter[i] counts the number of (logical) nodes retired by process $p_i$. If \inCounter[i] $>$ \outCounter[i], then it implies that $p_i$ has an active request. Each process takes a snapshot of the \inCounter[] counter of all processes in an incremental manner and then waits for the \outCounter[] counter to ``catch up''. After $2n$ requests, it swaps its active pool with its reserve pool. After $4n$ requests, nodes get so old that no other process has any reference to them. Thus, it is safe to allow node allocation for process $p_i$ repeats after it has finished $4n$ requests.
            
            Note that the pseudocode in \autoref{algo:memrec} is for the CC model. A similar memory reclamation algorithm can be implemented for the DSM model using a notification based system.

        \subsection{Conclusion}
            In this work, we have described a general framework to transform any non-adaptive RME algorithm into a \sadaptive{} one without increasing its worst-case RMR complexity. In addition to the hardware instructions used by the underlying non-adaptive RME algorithm, our framework uses \CAS{} and \FAS{} RMW instructions, both of which are commonly available on most modern processors.  
            When applied to the non-adaptive RME algorithm proposed by \JJJ{} in~\cite{JayJay+:2019:PODC}, it yields a \sefficient{} RME algorithm whose RMR complexity is $\bigO{\min\{ F_b + \sqrt{\F}, \nicefrac{\log \n}{\log \log \n} \}}$. 
        
            In this work, a failed process, upon restarting, attempts to reacquire all the locks at every level, beginning from level one. As a result, the worst case RMR complexity of the super-passage of such a process is $\bigO{\F_0 * \min \{ \sqrt{\F}, \nicefrac{\log \n}{\log \log \n} \}}$, where $\F_0$ denotes the number of times the process fails while executing its super-passage. However, if a process keeps track of its last known level, the worst case RMR complexity of the super passage of such a process can be reduced to $\bigO{\F_0 + \min \{ \sqrt{\F}, \nicefrac{\log \n}{\log \log \n} \}}$.

\bibliographystyle{ACM-Reference-Format}
\bibliography{Citations,References}

\end{document}